\documentclass[a4paper,fleqn]{cas-dc}

\usepackage[numbers,sort&compress]{natbib}
\usepackage{amsmath,amssymb}
\usepackage{xcolor}
\usepackage{url}
\usepackage{algorithm}
\usepackage{paralist}
\usepackage{xspace}
\usepackage{graphicx}
\usepackage{tabularx}
\usepackage{subcaption}
\usepackage{multirow}
\usepackage{booktabs}
\usepackage{enumitem}
\usepackage{algpseudocode}
\usepackage{tikz}
\usepackage{float}
\usetikzlibrary{positioning, arrows.meta, shapes.geometric}
\usetikzlibrary{decorations.pathmorphing}

\newcommand{\fakeparagraph}[1]{\vspace{1mm}\noindent\textit{#1.}}

\newcommand{\fig}[1]{Fig.~\ref{fig:#1}}
 \newcommand{\s}[1]{\S\ref{sec:#1}}

\newcommand{\setN}{\mathcal{N}} \newcommand{\setF}{\mathcal{F}}
 \newcommand{\setC}{\mathcal{C}}
\newcommand{\setI}{\mathcal{I}}

\newcommand{\CSR}{\setC^{\mathrm{SR}}} \newcommand{\CER}{\setC^{\mathrm{ER}}}
\newcommand{\CfSR}{\setC_f^{\mathrm{SR}}}
\newcommand{\CfER}{\setC_f^{\mathrm{ER}}} \newcommand{\Fr}{\setF_{\mathrm{r}}}
\newcommand{\Fw}{\setF_{\mathrm{w}}} \newcommand{\texec}{t^{\mathrm{exec}}}
\newcommand{\tretr}{t^{\mathrm{retr}}} \newcommand{\Texec}{T_{\mathrm{exec}}}
\newcommand{\Tretr}{T_{\mathrm{retr}}}

\algnewcommand\Phase[1]{\Statex \textbf{#1 Phase:}}

\newif\ifhidenotes
\hidenotestrue

\ifhidenotes
  \newcommand{\noteam}[1]{}
  \newcommand{\notedd}[1]{}
  \newcommand{\notead}[1]{}
  \newcommand{\notemc}[1]{}
\else
  \newcommand{\noteam}[1]{\footnote{\color{red}{Ale: #1}}}
  \newcommand{\notedd}[1]{\footnote{\color{blue}{Dario: #1}}}
  \newcommand{\notead}[1]{\footnote{\color{purple}{Ari: #1}}}
  \newcommand{\notemc}[1]{\footnote{\color{teal}{Matteo: #1}}}
\fi

\newif\ifanonymous
\anonymousfalse

\def\tsc#1{\csdef{#1}{\textsc{\lowercase{#1}}\xspace}}
\tsc{WGM}
\tsc{QE}

\begin{document}
\let\WriteBookmarks\relax \def\floatpagefraction{.99} \def\textfraction{.001}

\shorttitle{Replication Meets Scheduling}
\ifanonymous
  \shortauthors{Anonymous Author(s)}
\else
  \shortauthors{Cenzato et al.}
\fi

\title [mode = title]{Data Replication Meets Function Scheduling in the Edge-Cloud Continuum}

\ifanonymous
  \author[1]{Anonymous Author(s)}
  \affiliation[1]{organization={Anonymous Organization}, city={Anonymous
        City}, country={Anonymous Country}}
\else
  \author[1]{Matteo Cenzato}\ead{matteo.cenzato@mail.polimi.it}
  \credit{Conceptualization, Methodology, Writing - review and editing}

  \author[1]{Dario d'Abate}\cormark[1]\ead{dario.dabate@polimi.it}
  \credit{Conceptualization, Methodology, Software, Validation, Formal analysis, Investigation, Visualization, Writing - original draft, Writing - review and editing}

  \author[1]{Arianna Dragoni}\ead{arianna.dragoni@polimi.it}
  \credit{Conceptualization, Methodology, Supervision, Writing - review and editing}

  \author[1]{Matteo Briscini}\ead{matteo.briscini@mail.polimi.it}
  \credit{Conceptualization, Methodology}

  \author[1]{Alessandro Margara}\ead{alessandro.margara@polimi.it}
  \credit{Conceptualization, Formal analysis, Validation, Resources, Funding acquisition, Supervision, Writing - review and editing}

  \affiliation[1]{organization={Politecnico di Milano},
    addressline={Piazza Leonardo da Vinci 32}, postcode={20133}, city={Milano}, country={Italy}}

  \cortext[1]{Corresponding author}
\fi

\begin{abstract}
  Serverless computing is an appealing model for the edge-cloud continuum, but
  its stateless assumption breaks down once functions need persistent data:
  fetching state from a distant cloud store erases the latency benefit of
  running at the edge. Keeping data close means replicating it, and
  replication forces a placement decision that is coupled with where functions
  execute and with the consistency each application demands. We study this
  joint problem of function scheduling and data placement under two
  consistency models, strong and eventual replication. We first formulate it
  as a Binary Linear Program that yields the optimal placement for a given
  system snapshot, and use it as a reference point. Because the solver does
  not scale past a few hundred nodes, we add two heuristics with progressively
  less information: a Global-View greedy method that works from the same
  complete snapshot, and an Aggregated-View heuristic in which each node
  decides from locally observed demand alone. Across a range of system sizes
  the Global-View heuristic stays within a few percent of the optimum while
  scaling to over $10^4$ nodes. The Aggregated-View heuristic sacrifices some
  solution quality, but adapts continuously to each invocation. Taken
  together, the results show that the most suitable decision mechanism depends
  on workload stability: under client mobility, centralized policies suffer
  from stale snapshots and recurring latency spikes, while the Aggregated-View
  maintains low and stable client-observed latency. Across all experiments,
  data placement proves more influential than function scheduling in
  determining the outcome.
\end{abstract} 


\maketitle

\section{Introduction}
\label{sec:intro}

Cloud computing has long been the de facto target for deploying
applications~\cite{gannon:CloudComp:2017:cloud-native}, offering managed
environments with virtually unlimited resources. However, in recent years, the
emergence of the edge-cloud continuum has fragmented the landscape of
computing infrastructure~\cite{bittencourt:continuum:2025:compsci-review}.
Edge devices, deployed close to end users, enable lower latencies for
applications that are sensitive to response
times~\cite{shi:edge-comp:iot-journal:2016}. This is particularly relevant for
domains such as autonomous driving and industrial IoT, where systems operate
under strict low-latency constraints.

One of the dominant programming paradigms proposed for the edge-cloud
continuum, especially in academic proposals, is serverless
computing~\cite{nastic:serverless-edgetocloud:internet-comp:2017}, and in
particular its Function-as-a-Service (FaaS) model, where developers decompose
their applications into functions, and managed runtimes handle their scaling
and placement across the available nodes.
The FaaS model is inherently stateless~\cite{jonas:serverless:2019EECS}: in
cloud environments, when applications require persistent state, this is
delegated to external data stores co-located in the same data
center~\cite{jonas:serverless:2019EECS}.
When functions execute at the edge, accessing a remote cloud-hosted store
incurs prohibitive latency, negating the benefits of edge
deployment~\cite{shi:edge-comp:iot-journal:2016}.

Removing this limitation gives rise to a challenging optimization problem. In
stateful serverless, we must decide not only where to execute each function
(i.e., \emph{function scheduling} problem), but also where to place the data
items that functions access (i.e., \emph{data placement} problem).
Since multiple functions may access the same data items from different nodes,
the placement of each data item involves a trade-off: placing it close to one
consumer may penalize another. We refer to the combination of these two
decisions as the \emph{placement problem}.

A natural approach to mitigate this trade-off is \emph{replication}: placing
copies of data items on multiple nodes so that each function can access a
nearby replica. However, replication introduces its own challenges. Additional
replicas reduce read latency by bringing data closer to consumers, but updates
must be propagated across all copies.

Furthermore, different applications impose different \emph{consistency
  requirements} on their data, ranging from strong guarantees that synchronize
all replicas before serving any read, to weaker guarantees that allow
replicas to diverge temporarily and converge over time. A placement
framework must therefore account for these consistency semantics, as they
fundamentally affect both the feasible placement space and the resulting
latency.

Existing approaches to function scheduling and data placement can be
characterized along two dimensions. The first concerns \emph{what is modeled}:
some works focus exclusively on function scheduling, ignoring the data
dependencies of the functions they schedule~\cite{vahabi:energyEfficient};
others consider data placement but treat consistency coarsely, without
distinguishing among different consistency
models~\cite{nardelli:functionOffloading}.

The second dimension concerns \emph{how decisions are made}. Centralized
approaches based on exact solvers~\cite{baresi:neptune} model the optimum but
incur prohibitive computational overheads as the system grows. Faster
alternatives reduce this cost through
heuristics~\cite{rausch:containerScheduling}, which typically still rely on a
global view of the system, whose collection is itself expensive in the
edge-cloud continuum~\cite{ascigil:tsc:2022:faas-edge}. Decentralized
heuristics~\cite{nardelli:functionOffloading,cicconetti:faasInIoT} avoid this
requirement, at the price of operating on partial information.
What is currently missing in the literature is, first, a rigorous formulation
of the optimal placement baseline under heterogeneous consistency models,
second, a study of how closely heuristics with progressively narrower
information scope, from centralized to decentralized, can approximate this
baseline under realistic edge-cloud conditions, and third, an assessment of
whether the information advantage of a global view translates into better
adaptivity when demand shifts at runtime, or whether a decentralized approach
that reacts to each invocation individually can compensate for its narrower
view.

In this paper, we address this gap for the joint problem of function
scheduling and data placement in the edge-cloud continuum, under two
consistency models: strong replication (SR) and eventual replication (ER). Our
contributions are as follows.
We formalize the joint placement problem and propose a Binary Linear
Programming (BLP) model that jointly optimizes function scheduling and data
replica placement under SR and ER, serving as the theoretical benchmark
against which the other approaches are evaluated. We then propose a
\emph{Global-View} (GV) greedy heuristic that approximates the optimum
efficiently from the same complete snapshot, and an \emph{Aggregated-View}
(AV) heuristic that bypasses the dual overhead of system-wide information
collection and the subsequent scattering of placement decisions, each node
operating on a local, partial view of the system state.
Our evaluation shows that
\begin{inparaenum}[(i)]
  \item GV closely tracks the BLP optimum at a fraction of its computational
  cost, scaling to system sizes the solver cannot handle, while AV produces
  competitive latency through reactive replication at the cost of higher
  storage;
  \item in realistic deployments, AV bypasses the gather-scatter overhead of
  centralized approaches while delivering competitive client-observed latency;
  and
  \item under shifting access patterns, AV's per-invocation reactivity adapts
  continuously, avoiding the re-planning cost of centralized snapshots that
  quickly grow stale.
\end{inparaenum}

The rest of the paper is organized as follows.
\s{related} reviews related work.
\s{system_model} formalizes the joint placement problem.
\s{centralized_mod} presents the BLP model.
\s{topology_aware_heuristic} describes GV.
\s{decentralized_mod} introduces AV.
\s{eval} reports the empirical evaluation.
\s{conclusions} concludes the paper.
\section{Related Work}
\label{sec:related}

We organize this section along the two dimensions of \s{intro}: what existing
approaches model, and how they make decisions. First, however, a note on
scope: some works~\cite{xu:statefulServerless,yu:followingData} use
\emph{stateful serverless} to refer to function chaining, where ephemeral data
flows between stages of a workflow. We instead consider functions that access
\emph{persistent shared state} surviving across invocations.

\fakeparagraph{Function scheduling without data placement}
Several works optimize where to execute functions in the edge-cloud continuum
without modeling their data dependencies. Cicconetti et
al.~\cite{cicconetti:faasInIoT} address dynamic scheduling and offloading in
IoT environments: each edge router decides autonomously where to send
requests, using weighted averages of response times and information from an
SDN controller. Vahabi et al.~\cite{vahabi:energyEfficient} formulate an
Integer Linear Programming (ILP) problem that minimizes energy consumption for
stateless function scheduling at the edge; the model requires global knowledge
of all node capacities and power profiles, and is resolved using a centralized
solver. Baresi et al.~\cite{baresi:neptune} formulate a Mixed Integer Linear
Programming (MILP) for serverless placement on Multi-access Edge Computing
(MEC) infrastructure, where the topology is partitioned into communities and
the problem is solved centrally within each one; the access to external data
stores is not a decision variable but only a non-controllable component of the
response time. Rausch et al.~\cite{rausch:containerScheduling} extend the
Kubernetes scheduler with a scoring function that includes data proximity
among its criteria, but such proximity requires manual annotations from the
developer, and data is not placed by the system.

None of these works models data placement explicitly. In the edge-cloud
continuum, where functions and their data may reside on different nodes,
neglecting this coupling can result in significant latency penalties.

\fakeparagraph{Joint function scheduling and data placement}
Other works recognize that function scheduling and data placement are coupled.
Nardelli et al.~\cite{nardelli:functionOffloading} jointly model function
offloading and data migration: the offloading part is decentralized, with each
node deciding based on local knowledge, while data migration assumes a
logically centralized migrator. They explicitly leave replication out of their
model. Puliafito et al.~\cite{puliafito:stateAllocation} formulate a
centralized MILP that separates the problem into two sequential steps: first
allocating stateful microservices whose state is bound to the container, then
dispatching invocations to stateless serverless functions whose state remains
in the cloud. For the serverless part, data is not placed and replication is
not considered. Smith et al.~\cite{smith:fado} build a platform that routes
invocations to clusters holding the required data buckets and supports
replication across clusters via asynchronous mirroring (MinIO), but the
replication decisions are defined manually by the operator and no optimization
model guides the placement.

These works couple functions and data in the placement decision, but they
either exclude replication or do not distinguish among different consistency
semantics of the replicated data.

\fakeparagraph{Consistency in stateful serverless}
A few systems address consistency, but at the protocol level rather than as
part of the placement model. Cloudburst~\cite{sreekanti:cloudburst} is a
stateful FaaS platform built on Anna~\cite{wu:ICDE:2018:anna-kvs}, a
lattice-based distributed key-value store. Functions and data are co-located
through local mutable caches, and replicas emerge implicitly from back
pressure: when a node is overloaded, requests go to other nodes that fetch and
cache the data. The platform provides repeatable read and causal consistency
through distributed session protocols. While Cloudburst provides system-level
consistency mechanisms, it does not incorporate consistency constraints as a
variable in the placement decision. Pfandzelter et
al.~\cite{pfandzelter:enoki,pfandzelter:fred} go further in terms of
consistency awareness. Their platform integrates FReD, a replication
middleware that uses optimistic replication with version vectors and offers
configurable client-centric guarantees (monotonic reads, read your writes,
monotonic writes, write follows reads) per keygroup. This makes it the closest
work to ours in terms of consistency support. However, the decision of where
to replicate each keygroup is made manually by the application through an API,
and data is replicated on all designated nodes regardless of access patterns
or costs. Lambdata~\cite{tang:lambdata} avoids the problem by design: data is
immutable, and a function that wants to modify a data item writes a new key
and deletes the old one, so cache consistency never arises.
LoLa~\cite{wen:lola} manages state migration at the edge with fine-grained
locking and read/write separation, but again consistency is an operational
mechanism, not a placement variable.

None of these systems integrates heterogeneous consistency semantics as a core
dimension of an automated, joint placement optimization. %
\section{System Model and Problem Statement}
\label{sec:system_model}

This section presents the system model used throughout the paper. The model
captures typical characteristics of edge-to-cloud environments, including a
hierarchical node infrastructure (\s{system_model:infrastructure}), stateful
serverless functions that access potentially shared data
(\s{system_model:application}), and heterogeneous data consistency strategies
(\s{system_model:consistency}). Based on these elements, we formalize the
placement problem (\s{system_model:problem}).

\subsection{Infrastructure}
\label{sec:system_model:infrastructure}

We consider a set $\setN = \{1, \dots, N\}$ of heterogeneous nodes spanning
the edge-cloud continuum. A node $j \in \setN$ is characterized by its
computational speed $\mathit{speed}_j$ and storage capacity $\mathit{stor}_j$.

Edge-to-cloud deployments are usually arranged in layered tiers:
conventionally \emph{Cloud}, \emph{Fog}, and \emph{Edge}, where each layer may
comprise several logical
levels~\cite{bittencourt:continuum:2025:compsci-review}.
Node density decreases from Edge to Cloud, and traffic typically aggregates
uplink toward the root; we therefore model the infrastructure as a
hierarchical tree where the Cloud is the root, a logical node with unlimited
resources, possibly representing a data-center cluster. The remaining nodes
form arbitrary subtrees rooted at the cloud, capturing deployments with any
number of intermediate tiers between Cloud and Edge.
Nodes are connected through the links of the tree. For every pair of nodes
$(i, j)$, we denote by $\mathit{lat}_{ij}$ the network latency between them,
computed as the sum of the latencies of the links along the unique path
connecting them in the tree, and by $\mathit{ban}_{ij}$ their effective
bandwidth, determined by the minimum bandwidth among the links on the same
path. We assume symmetric channels, so $\mathit{lat}_{ij} = \mathit{lat}_{ji}$
and $\mathit{ban}_{ij} = \mathit{ban}_{ji}$. Data retrieval from local storage
is considered negligible with respect to network latency.

\subsection{Application Model}
\label{sec:system_model:application}

Applications are composed of stateful serverless functions.
In the FaaS model, functions have a two-level structure: a \emph{function
  class} defines the code and resource requirements, and is \emph{registered}
on the platform once; \emph{invocations} are individual executions of that
class, triggered by client requests.
Running a function on a node requires the class to be \emph{deployed} there
first, which means that its runtime must be active and its code available
locally.
When multiple invocations of the same function converge on the same node, a
single deployment serves them all.
In our model, we abstract away the resource cost of deployment and focus
exclusively on scheduling invocations and placing data. This is consistent
with the serverless model, where cold-start costs are amortized across many
invocations and the dominant latency component is data access.
Similarly, we do not model per-node computational resource limits (CPU,
memory), as the memory footprint of a function invocation depends on the
runtime input and is difficult to bound statically. The dominant cost in our
target scenario is data access latency, not compute contention.
Let $\setF = \{f_1, \dots, f_F\}$ denote the set of registered functions.
Each function $f$ is characterized by a reference execution time
$\mathit{ref}_f$.

End users (\emph{clients}) invoke functions from anywhere in the hierarchy;
the demand is captured by the invocation rate $\lambda_{f,i}$, which gives the
rate at which function $f$ is invoked from node $i$. We collect all pairs $(f,
  i)$ with $\lambda_{f,i} > 0$ in the set: $ \setI = \{(f, i) \in \setF \times
  \setN : \lambda_{f,i} > 0 \}$.
Scheduling is per pair $(f, i) \in \setI$, allowing invocations of the same
registered function from different sources to be routed independently; when
they converge on the same execution node, however, the function is deployed
there only once.

As data abstraction, we adopt the notion of \emph{collection}. A collection in
our model represents an opaque unit of state whose internal structure is
managed by the application or by an external data store sitting above our
placement layer.
Concretely, a collection may wrap a composite data structure (e.g., documents,
tables, or nested records). For instance, in a travel agency application, one
collection might aggregate a customer's flight bookings, hotel reservations,
and loyalty information, whereas a more normalized application would use
finer-grained collections. Both designs are valid: what matters for the
placement model is the collection as the atomic unit of replication, not its
internal layout.

Let $\setC = \{c_1, \dots, c_C\}$ denote the set of collections. Each
collection $c$ has a known size $\mathit{size}_c$ and a consistency strategy
$\mathit{sr}_c$, where $\mathit{sr}_c = 1$ denotes strong replication (SR) and
$\mathit{sr}_c = 0$ denotes eventual replication (ER).

Functions access the collections they require either locally, when co-located
on the same node, or remotely, by fetching data from another node at the cost
of additional network latency. The set of collections accessed by each
function, as well as the type of access (read or write), must be known
statically for a given problem instance; in practice, this information can be
derived from static annotations provided by the
developer~\cite{rausch:containerScheduling} or automated code
analysis~\cite{li:r3:vldb:2023}. The access relation is modeled through the
parameter $\mathit{access}_{c,f}$; we write $\setC_f = \{c \in \setC :
  \mathit{access}_{c,f} = 1\}$ for the set of collections accessed by function
$f$. A function is classified as \emph{read-only} ($\mathit{read}_f = 1$) if
all its accesses are reads, and as a \emph{writing function} ($\mathit{read}_f
  = 0$) if it performs at least a write access.

\subsection{Consistency Strategies}
\label{sec:system_model:consistency}

Even within a single application, different collections may require different
consistency guarantees.
In the travel agency above, a collection tracking available seat counts must
reflect every booking immediately, with all replicas agreeing before a new
request is served to prevent the flight from being overbooked. A collection
aggregating total searches per destination, by contrast, tolerates temporarily
divergent replicas: missing one update violates no invariant, and the value
converges once updates propagate.
Accordingly, each collection is associated with one of two consistency
strategies, which determine how it can be replicated and accessed: we refer to
them as \emph{SR collections} ($\mathit{sr}_c = 1$) and \emph{ER collections}
($\mathit{sr}_c = 0$).

\fakeparagraph{Strong Replication (SR)}
Under SR, collections can be replicated across multiple nodes while preserving
a total order on write operations.
We model this guarantee through a single-leader
protocol~\cite{kleppmann:designingDataIntensive}: one replica per collection
is designated as the \emph{leader} and serves all writes, while any replica
can serve reads.
Single-leader replication is the most widely adopted design for strongly
consistent
data\footnote{\url{https://etcd.io}}\footnote{\url{https://tikv.org}} because
it avoids the per-operation quorum overhead that characterizes multi-leader
and leaderless alternatives, an overhead we seek to avoid in edge-to-cloud
deployments. The leader propagates updates to followers asynchronously,
outside the critical path of function invocations. The total order of updates
is preserved because the unique path connecting any two nodes in the tree
consists of FIFO communication channels, consistent with standard reliable
networking assumptions in state
replication~\cite{kleppmann:designingDataIntensive}.
We impose two constraints on functions that access SR collections. First, each
function accesses at most one SR collection (alongside any number of ER
collections). Second, a function that accesses an SR collection must execute
on a node that holds a replica of that collection; specifically, a writing
function must be co-located with the \emph{leader} replica, so that write
operations do not incur an additional remote round-trip, while a read-only
function requires any local replica.

Both constraints are motivated by the cost of cross-collection coordination in
geo-distributed environments. Distributed databases supporting multiple
consistency levels face the problem of determining the scope of transactional
guarantees, and many restrict atomicity to avoid global coordination
bottlenecks: Apache Cassandra\footnote{\url{https://cassandra.apache.org}},
for instance, provides Lightweight Transactions (LWT) that guarantee
linearizability only within a single partition, and no cross-partition
transaction. More generally, strong consistency across independent,
potentially geo-distributed collections requires global coordination (e.g.,
two-phase commit), whose stop-the-world barrier blocks every participant until
all have voted, an expense that is prohibitive when inter-node latencies are
orders of magnitude larger than within a datacenter. Restricting each function
to a single SR collection removes cross-collection coordination entirely, and
co-location removes the synchronous remote round-trip from the critical path
of each invocation.

\fakeparagraph{Eventual Replication (ER)}
Under ER, collections can be replicated without a designated leader. Any
replica can serve both reads and writes, with updates propagating
asynchronously and conflicts resolved through application-level policies
(e.g., last-writer-wins, CRDTs~\cite{shapiro:crdt:2011}).

Functions are free to access multiple ER collections within a single
invocation, as conflicts will eventually be solved.
Likewise, we relax the co-location requirement: a function may execute without
local copies of all its ER collections, fetching missing ones from remote
replicas at the cost of additional retrieval latency.

These two strategies induce a \emph{structural asymmetry} in how an added
replica helps. For an ER collection, every additional replica is an
independent, additive benefit: a copy near a consumer cuts its retrieval
latency without affecting any other collection, so the only opposing factor is
node capacity. For an SR collection, the single-leader constraint pins all
writes to one contended location; an added replica is therefore a read-only
follower that shortens only read paths, while the write workload stays bound
to the leader. This asymmetry recurs in every placement approach that follows.

\fakeparagraph{Fault tolerance}
Across both strategies, at least one copy of each collection must exist at all
times. Our model assumes that a copy of each collection always persists in the
cloud, where storage is abundant, providing a baseline for disaster recovery.
Beyond this assumption, fault tolerance is outside the scope of this work: we
do not model node failures, leader re-election, or replica recovery, and focus
exclusively on the placement problem under normal operating conditions.

\subsection{Placement Problem}
\label{sec:system_model:problem}

Given the infrastructure, the set of registered functions with their
invocation statistics, and the set of collections with their consistency
strategies and access relations, the \emph{placement problem} consists of two
coupled decisions: (1)~\emph{data placement}: on which nodes to place each
collection and its replicas, and
(2)~\emph{invocation scheduling}: on which node to execute each function
invocation originated from a source node.

These decisions are tightly connected: where collections are placed constrains
where functions can run, and where functions run determines the cost of
accessing remote collections (under ER).

The primary goal is to minimize the expected invocation latency, which
comprises the network delay between the client and the execution node, the
function execution time, and, under ER, the cost of fetching collections that
are not locally available. At the same time, replication must be kept under
control: placing replicas on every node would trivially eliminate remote
access latency, but at the cost of saturating the limited storage of edge
nodes. The placement must therefore balance data proximity against resource
consumption, subject to the finite storage capacity of each node and the
co-location requirements imposed by the chosen consistency strategy.

\section{Centralized Model}
\label{sec:centralized_mod}

This section formalizes the placement problem (\s{system_model}) as a BLP. Our
formulation handles both SR and ER collections: the $\mathit{sr}_c$ parameter
activates the appropriate constraints and latency terms. For convenience, we
partition the collections accessed by function $f$ into $\CfSR = \{c \in
  \setC_f : \mathit{sr}_c = 1\}$ and $\CfER = \{c \in \setC_f : \mathit{sr}_c =
  0\}$, and define the global sets $\CSR$ and $\CER$ analogously. As discussed
in \s{system_model:consistency}, we require $|\CfSR| \leq 1$ for every $f$.
Similarly, we partition functions into read-only $\Fr = \{f \in \setF :
  \mathit{read}_f = 1\}$ and writing $\Fw = \{f \in \setF : \mathit{read}_f =
  0\}$.

\subsection{Decision Variables and Auxiliary Indicators}
\label{sec:centralized_mod:vars}

The formulation uses four families of primary binary decision variables, which
encode the placement, scheduling, leader election, and routing choices.

The \emph{placement variable} $x_{c,j} = 1$ iff a replica of collection $c$ is
placed on node $j$. The \emph{scheduling variable} $w_{f,i,j} = 1$ iff the
invocation of function $f$ from source $i$ is scheduled on node $j$. The
\emph{leader election variable} $l_{c,j} = 1$ iff node $j$ holds the leader
replica of SR collection $c$. The \emph{routing variable} $r_{f,i,c,j,a} = 1$
iff the invocation of $f$ from $i$, executing on $j$, fetches the missing ER
collection $c$ from provider node $a$.

  {\footnotesize
    \begin{align}
      x_{c,j}       & \in \{0,1\} & \forall\, & c \in \setC,\, j \in \setN \label{eq:var-x}      \\
      w_{f,i,j}     & \in \{0,1\} & \forall\, & (f,i) \in \setI,\, j \in \setN \label{eq:var-w}  \\
      l_{c,j}       & \in \{0,1\} & \forall\, & c \in \CSR,\, j \in \setN \label{eq:var-l}       \\
      r_{f,i,c,j,a} & \in \{0,1\} & \forall\, & \begin{aligned}[t]
                                                  (f,i) \in \setI,\, c \in \CfER,\, j \in \setN, \\
                                                  a \in \setN \setminus \{j\}
                                                \end{aligned} \label{eq:var-r}
    \end{align}
  }

One family of auxiliary binary decision variables (hereafter \emph{auxiliary
  indicators}) is used to linearize logical dependencies.

The \emph{miss indicator} $z_{f,i,c,j} = 1$ iff the invocation of $f$ from $i$
is scheduled on $j$ and ER collection $c$ is not locally available. This
indicator captures the nonlinear product $w_{f,i,j} \cdot (1 - x_{c,j})$,
linearized in \s{centralized_mod:constraints}.

  {\footnotesize
    \begin{align}
      z_{f,i,c,j} & \in \{0,1\} & \forall\, & (f,i) \in \setI,\, c \in \CfER,\, j
      \in \setN
      \label{eq:var-z}
    \end{align}}

\subsection{Objective Function}
\label{sec:centralized_mod:obj}

The placement objective minimizes a weighted combination of the rate-weighted
total latency $T$ and the total storage footprint $S$ \eqref{eq:weighted-obj}.
Incorporating storage directly into the objective reflects the cost of
replication. A formulation that minimizes latency alone is free to saturate
every node's storage as long as doing so yields any reduction, however
marginal.
The parameter $\alpha \geq 0$ governs this latency--storage trade-off: at
$\alpha = 0$ the model replicates as aggressively as node capacity allows; as
$\alpha$ grows, the cost of each additional replica is weighed more heavily,
producing sparser placements.
Both terms are normalized to the $[0, 1]$ range using their theoretical
worst-case bounds, $T^{\max}$ and $S^{\max}$. The rate-weighted total latency
$T$ decomposes into an execution component $\Texec$ and a retrieval component
$\Tretr$, both linear in the decision variables $w$ and $r$:

{
  \footnotesize
  \begin{align}
    \min \left( \frac{T}{T^{\max}} + \alpha \cdot \frac{S}{S^{\max}} \right),
    \quad T = \Texec + \Tretr.
    \label{eq:weighted-obj}
  \end{align}
}

We assume that invocation rates remain stationary over the optimization
horizon. This is consistent with the timescale of placement decisions, which
typically target aggregate traffic patterns rather than instantaneous
fluctuations. Since the total invocation rate $\Lambda = \sum_{(f,i) \in
    \setI} \lambda_{f,i}$ is a positive constant for a given problem instance,
minimizing the average latency $T / \Lambda$ is equivalent to minimizing the
rate-weighted total~$T$.

\noindent
$\Texec$ aggregates per-invocation execution latency, weighted by the
corresponding rate $\lambda_{f,i}$ \eqref{eq:t-exec-agg}. The per-invocation
latency $\texec_{f,i,j}$ combines the round-trip network latency between
source node $i$ and execution node $j$ with the on-node computation time
\eqref{eq:t-exec-pair}. Bandwidth is omitted: invocation payloads are assumed
negligible compared to the collection data transferred during retrieval.

  {\footnotesize
    \begin{align}
      \Texec         & = \sum_{(f,i) \in \setI} \sum_{j \in \setN}
      \lambda_{f,i} \cdot \texec_{f,i,j} \cdot w_{f,i,j}
      \label{eq:t-exec-agg}                                        \\
      \texec_{f,i,j} & = 2 \cdot \mathit{lat}_{i,j} +
      \frac{\mathit{ref}_f}{\mathit{speed}_j} \quad \forall\, (f,i) \in \setI,\,
      \forall\, j \in \setN
      \label{eq:t-exec-pair}
    \end{align}}

\noindent
$\Tretr$ aggregates per-fetch retrieval latency, again weighted by
$\lambda_{f,i}$ \eqref{eq:t-retr-agg}. The per-fetch cost is the round-trip
latency plus the transfer time of the collection over the available bandwidth,
and depends only on the link $(j,a)$ once $j$ is fixed \eqref{eq:t-retr-pair}.
This component is non-zero only for ER accesses: SR collections are guaranteed
to be locally available by the co-location constraints
(\s{centralized_mod:constraints}).

  {\footnotesize
    \begin{align}
      \Tretr         & = \sum_{(f,i) \in \setI} \sum_{c \in \CfER}
      \sum_{j \in \setN} \sum_{\substack{a \in \setN \\ a \neq j}} \lambda_{f,i}
      \cdot \tretr_{c,j,a} \cdot r_{f,i,c,j,a}
      \label{eq:t-retr-agg}                                        \\
      \tretr_{c,j,a} & = 2 \cdot \mathit{lat}_{j,a} +
      \frac{\mathit{size}_c}{\mathit{ban}_{j,a}} \quad \forall\, c \in \CER,\,
      j \in \setN,\, a \neq j
      \label{eq:t-retr-pair}
    \end{align}}

The worst-case rate-weighted latency $T^{\max}$ \eqref{eq:t-max} assumes, for
each $(f,i)$, the most expensive execution node and, for each ER collection,
the most expensive retrieval source.
The total storage cost $S$ \eqref{eq:obj-S} aggregates the size of all
replicas across all nodes, and its worst-case bound $S^{\max}$ corresponds to
replicating every collection on every node \eqref{eq:s-max}.

  {\footnotesize
    \begin{align}
      T^{\max} & = \sum_{(f,i) \in \setI} \lambda_{f,i} \cdot \max_{j \in \setN}
      \left( \texec_{f,i,j} + \sum_{c \in \CfER} \max_{\substack{a \in \setN \\ a
                                                     \neq j}} \tretr_{c,j,a} \right)
      \label{eq:t-max}                                                           \\
      S        & = \sum_{c \in \setC} \sum_{j \in \setN} \mathit{size}_c \cdot
      x_{c,j}
      \label{eq:obj-S}                                                           \\
      S^{\max} & = |\setN| \cdot \sum_{c \in \setC} \mathit{size}_c
      \label{eq:s-max}
    \end{align}}

\subsection{Constraints}
\label{sec:centralized_mod:constraints}

The constraints fall into four groups: structural constraints that apply to
all collections, SR-specific constraints, ER-specific constraints, and
resource constraints.

\fakeparagraph{Structural constraints}
Each invocation is scheduled on exactly one node
\eqref{eq:invocation-uniqueness}; every collection has at least one replica
\eqref{eq:collection-existence}.

  {\footnotesize
    \begin{align}
      \sum_{j \in \setN} w_{f,i,j} & = 1    & \forall\, & (f,i) \in \setI
      \label{eq:invocation-uniqueness}                                    \\
      \sum_{j \in \setN} x_{c,j}   & \geq 1 & \forall\, & c \in \setC
      \label{eq:collection-existence}
    \end{align}}

\fakeparagraph{SR-specific constraints}
Each SR collection has exactly one leader \eqref{eq:leader-uniqueness},
co-located with a replica \eqref{eq:leader-implies-replica}. Read-only
invocations must execute where a replica exists \eqref{eq:sr-colocation-read};
writing invocations must execute on the leader \eqref{eq:sr-colocation-write}.

  {\footnotesize
    \begin{align}
      \sum_{j \in \setN} l_{c,j} & = 1          & \forall\, & c \in \CSR
      \label{eq:leader-uniqueness}                                            \\
      l_{c,j}                    & \leq x_{c,j} & \forall\, & c \in \CSR,\,
      j \in \setN
      \label{eq:leader-implies-replica}                                       \\
      w_{f,i,j}                  & \leq x_{c,j} & \forall\, & (f,i) \in \setI
      \text{ with } f \in \Fr,\, c \in \CfSR,\, j \in \setN
      \label{eq:sr-colocation-read}                                           \\
      w_{f,i,j}                  & \leq l_{c,j} & \forall\, & (f,i) \in \setI
      \text{ with } f \in \Fw,\, c \in \CfSR,\, j \in \setN
      \label{eq:sr-colocation-write}
    \end{align}}

\fakeparagraph{ER-specific constraints}
The miss indicator $z_{f,i,c,j}$ captures the product $w_{f,i,j} \cdot (1 -
  x_{c,j})$. Since all variables are binary, we linearize it using standard
McCormick envelopes~\cite{mccormick:computability}
\eqref{eq:mc1}--\eqref{eq:mc3}. On a miss, exactly one provider node must be
selected \eqref{eq:routing-activation}, and only nodes holding a replica can
serve as providers \eqref{eq:routing-replica}.

  {\footnotesize
    \begin{align}
      z_{f,i,c,j}                                             & \leq w_{f,i,j}
      \label{eq:mc1}                                                                     \\
      z_{f,i,c,j}                                             & \leq 1 - x_{c,j}
      \label{eq:mc2}                                                                     \\
      z_{f,i,c,j}                                             & \geq w_{f,i,j} - x_{c,j}
      \label{eq:mc3}                                                                     \\
      \sum_{\substack{a \in \setN \\ a \neq j}} r_{f,i,c,j,a} & = z_{f,i,c,j}
      \label{eq:routing-activation}                                                      \\
      r_{f,i,c,j,a}                                           & \leq x_{c,a}
      \label{eq:routing-replica}
    \end{align}}
\noindent
for all $(f,i) \in \setI$, $c \in \CfER$, and $j \in \setN$, with $a \in
  \setN \setminus \{j\}$ in \eqref{eq:routing-replica}.

\fakeparagraph{Resource constraints}
The total size of replicas stored on a node cannot exceed its storage limit
\eqref{eq:storage-capacity}.

  {\footnotesize
    \begin{align}
      \sum_{c \in \setC} \mathit{size}_c \cdot x_{c,j} & \leq \mathit{stor}_j & \forall\, & j \in \setN
      \label{eq:storage-capacity}
    \end{align}}

\subsection{Model Size}
\label{sec:centralized_mod:mod_size}

All constraints and both objective components are linear in the decision
variables, making the formulation a BLP with a weighted-sum objective. The
routing variables $r_{f,i,c,j,a}$ have five indices and dominate the model
size with $O(|\setI| \cdot |\CER| \cdot |\setN|^2)$ entries in the worst case.
In practice, the number is significantly smaller: variables $z$ and $r$ are
only instantiated for function-collection pairs where $\mathit{access}_{c,f} =
  1$ and $\mathit{sr}_c = 0$, and each function typically accesses only a small
subset of collections. %
\section{Global-View Heuristic}\label{sec:topology_aware_heuristic}

While the BLP formulation of \s{centralized_mod} yields optimal solutions, its
variable count grows as $O(|\setI| \cdot |\setC^{\mathrm{ER}}| \cdot
  |\setN|^2)$, making it impractical for large instances or for periodic
reconfiguration. We propose a \emph{Global-View} (GV) greedy heuristic that
assumes a complete snapshot of the system state and exploits the hierarchical
tree infrastructure to approximate the optimum efficiently.

The algorithm operates in two phases. Phase~1 establishes a minimal feasible
baseline: it places a single replica of every collection, elects a leader for
each SR collection, and derives an initial scheduling and routing from the
resulting replica set. Phase~2 then greedily expands the replica set, adding
copies of both ER and SR collections wherever the latency savings outweigh the
storage cost, using the same $\alpha$ trade-off as the BLP objective
\eqref{eq:weighted-obj}.

This separation reflects the structural asymmetry between SR and ER
collections (\s{system_model:consistency}). Because an added ER replica
is an independent gain opposed only by capacity, and an added SR replica is a
read-only follower that leaves the Phase~1 leader untouched
\eqref{eq:sr-colocation-write}, each commit in Phase~2 is a locally optimal
step that cannot worsen the placement of any other collection, nor disturb
leader election.

Feasibility is maintained throughout by offloading excess storage demands up
the tree to the root, which we model as a cloud with virtually infinite
resources. Algorithm~\ref{alg:greedy} gives the complete procedure.

\begin{algorithm}[t]
  \footnotesize
  \caption{Global-View Heuristic}
  \label{alg:greedy}
  \begin{algorithmic}[1]
    \Statex \textbf{Phase 1: Minimal Baseline} \State $x \gets
      \textsc{PlaceCollections}()$ \State $l \gets \textsc{PickLeaders}(x)$
    \State $w \gets \textsc{ScheduleFunctions}(x, l)$ \State $r \gets
      \textsc{RouteFetches}(x, w)$

    \Statex \Statex \textbf{Phase 2: Greedy Expansion} \Repeat \ForAll{$c \in
        \setC$ \textbf{and} $j \in \setN$ \textbf{s.t.} $x_{c,j} = 0$} \State
    $j_{\mathrm{eff}} \gets \textsc{AncestorWithCapacity}(j,
      \mathit{size}_c)$ \If{$x_{c,j_{\mathrm{eff}}} = 1$} \State
    $\mathit{benefit}_{c,j} \gets -\infty$ \Else \State $\Delta L_{c,j}
      \gets \textsc{EstimateSavings}(c, j_{\mathrm{eff}}, w)$ \State
    $\mathit{benefit}_{c,j} \gets \Delta L_{c,j} - \alpha \cdot
      \mathit{size}_c$ \EndIf \EndFor \State $(c^\ast, j^\ast) \gets
      \arg\max_{c,j}\, \mathit{benefit}_{c,j}$
    \If{$\mathit{benefit}_{c^\ast, j^\ast} > 0$} \State
    $j^\ast_{\mathrm{eff}} \gets \textsc{AncestorWithCapacity}(j^\ast,
      \mathit{size}_{c^\ast})$ \State $x_{c^\ast, j^\ast_{\mathrm{eff}}}
      \gets 1$ \State $w \gets \textsc{UpdateScheduling}(w, c^\ast)$ \State
    $r \gets \textsc{UpdateRouting}(r, w)$ \EndIf
    \Until{$\mathit{benefit}_{c^\ast, j^\ast} \leq 0$}

    \Statex \State \Return $(x, l, w, r)$
  \end{algorithmic}
\end{algorithm}

\fakeparagraph{Collection placement}
Phase~1 places exactly one replica per collection to establish a feasible
baseline. The replica is placed on the node with the highest aggregate
\emph{access pressure}. For a node $i$, it is defined as the sum of invocation
rates $\lambda_{f,i}$ over all functions $f\in\setF$ that access $c$. The only
exception are SR collections that are subject to write operations. Because
their leader is pinned for the entire run, only write operations contribute to
the pressure ($f\in\Fw$). If the top candidate node lacks sufficient residual
storage, the replica is iteratively pushed up the tree toward the root until a
node with available capacity is found. Orphan collections with zero access
traffic are placed directly at the cloud (the tree root).

\fakeparagraph{Leader election}
Among the replicas of each SR collection $c \in \CSR$, we elect as leader the
one whose host node $i$ has the highest aggregate \emph{write pressure}
$\sum_{f\in\Fw:\,\mathit{access}_{c,f}=1}\lambda_{f,i}$, breaking ties by node
index for determinism. Since every write must execute on the leader, placing
it where the writers concentrate minimizes the latency of the write workload.
In the baseline each written SR collection has a single replica, so this step
merely confirms it as leader; the choice is made once and never revised, as
the followers added in Phase~2 carry only read traffic.

\fakeparagraph{Function scheduling}
Each invocation $(f,i)\in\setI$ is assigned an execution node according to
$f$'s consistency requirements. If $f$ accesses an SR collection, the
co-location constraints of \s{centralized_mod:constraints} fix the
target: the nearest replica for reads, or the leader for writes. If $f$
depends only on ER collections, the algorithm identifies the optimal execution
node to minimize the overall invocation cost (accounting for network RTT,
function execution time, and remote data fetch latency). It determines the
Lowest Common Ancestor (LCA) of the nearest ER replicas and the source node
$i$, and enumerates all candidate nodes on the paths spanning from the source
and the replicas to the LCA. The execution is assigned to the candidate node
that minimizes the total cost. Functions with no data dependency are scheduled
directly at the source $i$.

\fakeparagraph{Remote fetch routing}
For each ER access missing at the execution node, the provider is selected as
the nearest replica by hop count.

\begin{figure}
  \centering
  \begin{subfigure}{0.48\columnwidth}
    \centering
    \resizebox{!}{\linewidth}{%
      \begin{tikzpicture}[ nodec/.style={circle, draw, thick, minimum
              size=1.25cm, fill=white}, rep/.style={draw, rounded corners=2pt,
              inner sep=3pt, font=\small\sffamily, fill=gray!10},
          exec/.style={draw, rounded corners=2pt, inner sep=3pt,
              font=\small\sffamily\bfseries, fill=orange!20, draw=orange!60!black},
          tier/.style={font=\scriptsize\itshape, gray},
          fetch/.style={-{Stealth[scale=1]}, dashed, red!70!black, thick},
          invoke/.style={-{Stealth[scale=1]}, blue!55!black, thick},
          albl/.style={font=\scriptsize} ]
        \node[nodec] (cl) at (0, 2.7) {};
        \node[rep] (c2-cl) at ([yshift=0.3cm]cl.center) {$c_2$};
        \node[exec] (f-cl) at ([yshift=-0.3cm]cl.center) {$f$};
        \node[nodec] (fg) at (0, 1.35) {};
        \node[nodec] (ed) at (0, 0) {};
        \node[rep] (c1-ed) at (ed) {$c_1$};
        \node[tier, anchor=east] at ([xshift=-2pt]cl.west) {cloud};
        \node[tier, anchor=east] at ([xshift=-2pt]fg.west) {fog};
        \node[tier, anchor=east] at ([xshift=-2pt]ed.west) {edge};
        \node[font=\scriptsize\itshape, below=0pt of ed] {$i$};
        \draw[thick] (cl) -- (fg) -- (ed);
        \draw[fetch] (c1-ed.east) to[bend right=55]
        node[albl, pos=0.2, right=1pt, text=red!70!black] {}
        (f-cl.east);
        \draw[invoke] (ed.east) to[bend right=22]
        node[albl, pos=0.82, right=1pt, text=blue!55!black] {invoke}
        (cl.east);
      \end{tikzpicture}}
    \label{subfig:gv-before}
    \caption{}
  \end{subfigure}
  \hfill
  \begin{subfigure}{0.5\columnwidth}
    \centering
    \resizebox{!}{\linewidth}{%
      \begin{tikzpicture}[ nodec/.style={circle, draw, thick, minimum
              size=1.25cm, fill=white}, rep/.style={draw, rounded corners=2pt,
              inner sep=3pt, font=\small\sffamily, fill=gray!10},
          newrep/.style={draw, rounded corners=2pt, inner sep=3pt,
              thick, font=\small\sffamily, fill=green!15},
          exec/.style={draw, rounded corners=2pt, inner sep=3pt,
              font=\small\sffamily\bfseries, fill=orange!20, draw=orange!60!black},
          tier/.style={font=\scriptsize\itshape, gray},
          fetch/.style={-{Stealth[scale=1]}, dashed, red!70!black, thick},
          invoke/.style={-{Stealth[scale=1]}, blue!55!black, thick},
          albl/.style={font=\scriptsize} ]
        \node[nodec] (cl) at (0, 2.7) {};
        \node[rep] (c2-cl) at (cl) {$c_2$};
        \node[nodec] (fg) at (0, 1.35) {};
        \node[newrep] (c2-fg) at ([yshift=0.3cm]fg.center) {$c_2$};
        \node[exec] (f-fg) at ([yshift=-0.3cm]fg.center) {$f$};
        \node[nodec] (ed) at (0, 0) {};
        \node[rep] (c1-ed) at (ed) {$c_1$};
        \node[tier, anchor=east] at ([xshift=-2pt]cl.west) {cloud};
        \node[tier, anchor=east] at ([xshift=-2pt]fg.west) {fog};
        \node[tier, anchor=east] at ([xshift=-2pt]ed.west) {edge};
        \node[font=\scriptsize\itshape, below=0pt of ed] {$i$};
        \draw[thick] (cl) -- (fg) -- (ed);
        \draw[fetch] (c1-ed.east) to[bend right=42]
        node[albl, pos=0.28, right=1pt, text=red!70!black] {}
        (f-fg.east);
        \draw[invoke] (ed.east) to[bend right=16]
        node[albl, pos=0.72, right=1pt, text=blue!55!black] {invoke}
        (fg.east);
      \end{tikzpicture}}
    \label{subfig:gv-after}
    \caption{}
  \end{subfigure}
  \caption{One greedy-expansion iteration. Function $f$ accesses $c_1$ (on the
    edge) and $c_2$ (on the cloud); the client is at the edge ($i$).
    \textbf{(a)}~Execution is at the cloud; $c_1$ is fetched upward.
    \textbf{(b)}~After committing a replica of $c_2$ at the fog, rescheduling
    moves execution to the fog, which also shortens the $c_1$ fetch.}
  \label{fig:gv-loop}
\end{figure}
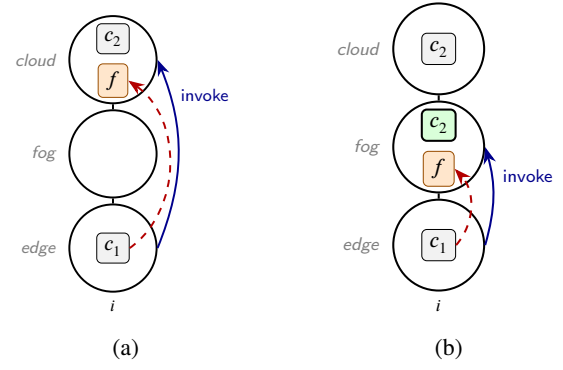

\fakeparagraph{Greedy expansion}
After the baseline, the algorithm iteratively adds replicas to reduce latency.
At each step, GV evaluates every candidate pair $(c, j)$ consisting of a
collection~$c$ and a node~$j$ that does not yet hold a replica of~$c$. For
each candidate, it estimates the rate-weighted latency reduction $\Delta
  L_{c,j}$ using the current scheduling as baseline. For SR, the estimate
predicts whether the new replica would become the nearest to the client,
yielding savings in $\texec$. For ER, re-evaluating all scheduling candidates
for each hypothetical replica would be too expensive, so the estimate
conservatively keeps the execution node fixed and evaluates savings in
$\tretr$ only. In both cases, only positive per-invocation savings are
accumulated, weighted by $\lambda_{f,i}$; SR writers are excluded. The net
benefit is $\Delta L_{c,j} - \alpha \cdot \mathit{size}_c$, mirroring the
trade-off in the BLP objective.
The candidate with the highest positive net benefit is committed: the replica
is placed, and all functions that access $c$ are rescheduled with their remote
fetches recomputed against the updated placement. This rescheduling can
trigger compounding gains that the per-candidate estimate does not anticipate:
a new replica may enable a closer execution node, which in turn shortens other
fetches (\fig{gv-loop}). The loop repeats until no candidate yields a positive
benefit.

\fakeparagraph{Complexity}
The baseline runs in $O(|\setI| \cdot |\setC| \cdot |\setN|)$. The greedy
expansion performs at most $|\setC| \cdot |\setN|$ iterations, each scanning
all $|\setC| \cdot |\setN|$ candidates, for a worst case of $O(|\setC|^2 \cdot
  |\setN|^2 \cdot |\setI|)$ if every candidate is tested against all
invocations. Our implementation instead pre-computes, for each collection, the
sparse set of \emph{active} invocations (those with $\lambda_{f,i} > 0$), so
\textsc{EstimateSavings} scans only the $K \ll |\setI|$ active
client-function pairs per collection, and the expansion costs $O(|\setC|^2
  \cdot |\setN|^2 \cdot K)$. It also halts early, as soon as the maximum
marginal benefit across all candidates drops to zero.

\section{Aggregated-View Heuristic} \label{sec:decentralized_mod}

Both the centralized BLP (\s{centralized_mod}) and the GV heuristic
(\s{topology_aware_heuristic}) start from a complete snapshot of the system
state, which is impractical to obtain at runtime in the edge-cloud continuum.
We therefore ask whether per-node decisions, taken over locally-available and
aggregated information alone, can track the same optimum without ever
materializing a global view.
Doing so requires decomposing the placement problem along the structural
asymmetry of \s{system_model:consistency}, addressing its two independent
sides with complementary mechanisms.
For ER, placement is a caching problem that each node drives reactively from
the requests it already serves (\s{er-dataPlacement}).
For SR, the contended leader's best position depends on write demand
aggregated across the tree that no node observes locally; we model this as a
force equilibrium that settles at the barycenter of write traffic, while reads
stay caching-like, served by followers placed near readers
(\s{sr-dataPlacement}).

The rest of the section describes the local state each node maintains
(\s{dec-state}) and the two runtime concerns it drives: data placement
(\s{dec-placement}) and function scheduling (\s{dec-scheduling}).
\fig{operational_state}(a) composes them into a single per-node loop, where
the local state is consulted by scheduling and updated by placement as traffic
is observed.

\subsection{Local State}
\label{sec:dec-state}

\begin{figure*}[htbp]
  \centering
  \begin{subfigure}{0.48\textwidth}
    \centering
    {\resizebox{0.35\columnwidth}{!}{%
        \begin{tikzpicture}[ stage/.style={draw, rectangle, rounded
                corners=3pt, thick, fill=white, minimum width=3.3cm, minimum
                height=1.05cm, align=center, font=\scriptsize},
            hub/.style={draw, rectangle, rounded corners=3pt, thick,
                fill=gray!12, minimum width=3.3cm, minimum height=1.05cm,
                align=center, font=\scriptsize},
            flow/.style={-{Stealth[scale=1.2]}, thick},
            ext/.style={-{Stealth[scale=1.2]}, thick, gray},
            lbl/.style={font=\tiny\itshape, align=center} ]

          \node[stage] (sched) at (0,  2.0) {\textbf{Function Scheduling}};
          \node[hub]   (state) at (0,  0)   {\textbf{Local State}\\
            $\rho_j,\,\ell_j,\,\mathcal{R}_j$, stats}; \node[stage] (place) at
          (0, -2.0) {\textbf{Data Placement}};

          \node[align=center, font=\scriptsize] (inv) at (0, 3.3) {Invocations
            $f$}; \draw[ext] (inv) -- (sched);

          \draw[flow] ([xshift=-0.7cm]state.north) --
            ([xshift=-0.7cm]sched.south) node[midway, left, lbl] {consult\\
              $\rho_j,\ell_j$}; \draw[flow] ([xshift=0.7cm]sched.south) --
            ([xshift=0.7cm]state.north) node[midway, right, lbl] {record\\
              traffic};

          \draw[flow] ([xshift=-0.7cm]state.south) --
            ([xshift=-0.7cm]place.north) node[midway, left, lbl] {read\\
              traffic}; \draw[flow] ([xshift=0.7cm]place.north) --
            ([xshift=0.7cm]state.south) node[midway, right, lbl] {update\\
              $\mathcal{R}_j$};

        \end{tikzpicture}%
      }}

    \caption{Per-node adaptation loop}
    \label{subfig:dynamic_pipeline}
  \end{subfigure}
  \hfill
  \begin{subfigure}{0.48\textwidth}
    \centering
    \resizebox{0.9\columnwidth}{!}{%
      \begin{tikzpicture}[
          node_circle/.style={circle, draw, minimum size=1.5cm, thick,
              fill=white}, replica_box/.style={draw, rounded corners=2pt,
              inner sep=2.5pt, font=\scriptsize\sffamily, fill=gray!10},
          table_node/.style={font=\scriptsize, fill=white, inner sep=2pt},
          route_arrow/.style={-{Stealth[scale=1.2]}, dashed, red, thick},
          tier_label/.style={font=\itshape\color{gray}, align=right} ]

        \node[node_circle] (n0) at (0, 0) {}; \node[node_circle] (n1) at
        (-1.53, -1.7) {}; \node[node_circle] (n2) at (1.53, -1.7) {};
        \node[node_circle] (n3) at (-2.38, -3.57) {};

        \node[right=2pt of n0] {$1$}; \node[right=2pt of n1] {$2$};
        \node[left=2pt of n2] {$3$}; \node[left=2pt of n3] {$4$};

        \draw[thick] (n0) -- (n1); \draw[thick] (n0) -- (n2); \draw[thick]
        (n1) -- (n3);

        \node[font=\scriptsize\itshape, gray] at ([yshift=0.5cm]n0.center)
        {$\mathcal{R}_1$}; \node[replica_box] at ([yshift=0.1cm]n0.center)
        {$c_1$}; \node[replica_box] at ([yshift=-0.35cm]n0.center) {$c_2$};

        \node[font=\scriptsize\itshape, gray] at ([yshift=0.35cm]n1.center)
        {$\mathcal{R}_2$}; \node[replica_box] at ([yshift=-0.15cm]n1.center)
        {$c_1$};

        \node[font=\scriptsize\itshape, gray] at ([yshift=0.35cm]n2.center)
        {$\mathcal{R}_3$}; \node[replica_box] (c2leader) at
        ([yshift=-0.15cm]n2.center) {$c_2$}; \node[font=\normalsize, right=1pt
          of c2leader, yshift=2pt] {$\bigstar$};

        \node[font=\scriptsize\itshape, gray] at ([yshift=0.35cm]n3.center)
        {$\mathcal{R}_4$}; \node[replica_box] at ([yshift=-0.15cm]n3.center)
        {$c_2$};

        \node[table_node, left=0.1cm of n0] {
          \setlength{\tabcolsep}{3pt}
          \begin{tabular}{@{}ccc@{}}
            \toprule
            $c$            & $\rho$                     & $\ell$ \\
            \midrule $c_1$ & $\{\mathsf{local}, 2\}$    & --     \\
            $c_2$          & $\{\mathsf{local}, 2, 3\}$ & $3$    \\
            \bottomrule
          \end{tabular}
        };

        \node[table_node, left=0.1cm of n1] {
          \setlength{\tabcolsep}{3pt}
          \begin{tabular}{@{}ccc@{}}
            \toprule
            $c$                  & $\rho$  & $\ell$        \\ \midrule $c_1$ &
            $\{\mathsf{local}\}$ & --                      \\
            $c_2$                & $\{4\}$ & $\mathsf{up}$ \\ \bottomrule
          \end{tabular}
        };

        \node[table_node, right=0.1cm of n2] {
          \setlength{\tabcolsep}{3pt}
          \begin{tabular}{@{}ccc@{}}
            \toprule
            $c$   & $\rho$               & $\ell$           \\ \midrule $c_1$
                  & $\{\mathsf{up}\}$    & --               \\
            $c_2$ & $\{\mathsf{local}\}$ & $\mathsf{local}$ \\ \bottomrule
          \end{tabular}
        };

        \node[table_node, right=0.2cm of n3] {
          \setlength{\tabcolsep}{3pt}
          \begin{tabular}{@{}ccc@{}}
            \toprule
            $c$            & $\rho$               & $\ell$        \\
            \midrule $c_1$ & $\{\mathsf{up}\}$    & --            \\
            $c_2$          & $\{\mathsf{local}\}$ & $\mathsf{up}$ \\
            \bottomrule
          \end{tabular}
        }; \end{tikzpicture}%
    }
    \caption{Local State}
    \label{subfig:static_state}
  \end{subfigure}

  \caption{Per-node local state (b) and its coupling with runtime operations
    (a). The replica store $\mathcal{R}_j$ holds the collections persisted on
    the node, while the reachability table ($\rho_j$, $\ell_j$) records the
    known directions to data. Scheduling queries this state; the resulting
    traffic metrics and data misses feed back into placement and eviction.}
  \label{fig:operational_state}
\end{figure*}
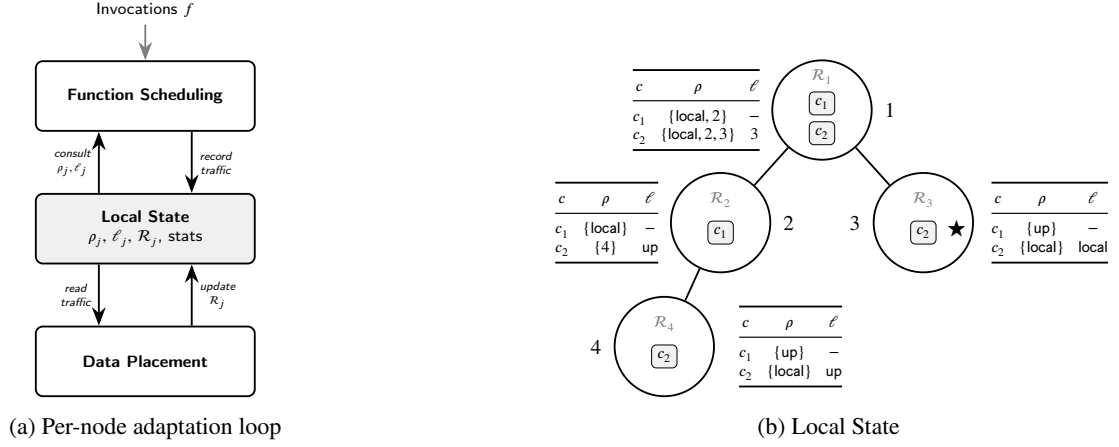

The heuristic operates on the same hierarchical infrastructure of
\s{system_model:infrastructure}, which we treat explicitly as a tree. For a
node $j \in \setN$, we write $\mathrm{par}(j)$ for its parent (undefined for
the root, i.e. the cloud), $\mathrm{ch}(j) \subseteq \setN$ for its set of
children, and $\mathrm{sub}(j) \subseteq \setN$ for the set of nodes in the
subtree rooted at $j$, with $j \in \mathrm{sub}(j)$. All node-local decisions
range over these three primitives only, never over the global node set
$\setN$.
Each node $j$ maintains a compact, incrementally updated \emph{local state},
sufficient to take scheduling and placement decisions, illustrated for a small
topology in \fig{operational_state}(b).

\fakeparagraph{Reachability table}
For every collection $c \in \setC$ it is aware of, node $j$ stores a set of
direction tags
\begin{align}
  \rho_j(c) \subseteq \{\mathsf{local}, \mathsf{up}\} \cup \mathrm{ch}(j),  \label{eq:reachability}
\end{align}
each marking a direction along which a replica of $c$ is known to be
reachable: $\mathsf{local}$ for a copy on $j$ itself, a child $k \in
  \mathrm{ch}(j)$ for a copy inside $k$'s subtree, and $\mathsf{up}$ when no
copy is known within $j$'s subtree, so $c$ is assumed reachable upstream,
toward the cloud, where the root always retains one
(\s{system_model:consistency}). A collection can be reachable in several
directions at once, both on $j$ itself and through more than one subtree,
which is why $\rho_j(c)$ is a set. For SR collections, where the leader is
unique, $j$ also tracks its single direction $\ell_j(c)$ under the same
semantics. Both are maintained incrementally as replicas are created or
migrated.

\fakeparagraph{Replica store}
Each node $j$ maintains a set $\mathcal{R}_j \subseteq \setC$ of collections
stored locally. For each collection $c \in \mathcal{R}_j$, $j$ tracks its
last-access timestamp $t_j^{\text{last}}(c)$ and a flag indicating whether $j$
hosts the leader for $c$. The latter implicitly defines a subset of local
leaders $\mathcal{L}_j \subseteq \mathcal{R}_j$. This replica store maps to
the reachability table: $\mathsf{local} \in \rho_j(c) \Leftrightarrow c \in
  \mathcal{R}_j$, and $\ell_j(c)=\mathsf{local} \Leftrightarrow c \in
  \mathcal{L}_j$. A node trivially derives its free storage capacity as
$\mathit{stor}^{\text{free}}_j = \mathit{stor}_j - \sum_{c \in \mathcal{R}_j}
  \mathit{size}_c$. When capacity is exhausted, the non-leader replicas
$\mathcal{R}_j \setminus \mathcal{L}_j$ are the candidates for eviction
(\s{dec-eviction}); leaders are exempt, being repositioned only by the
proactive SR protocol (\s{sr-dataPlacement}).

\fakeparagraph{Traffic statistics}
For each SR collection, $j$ accumulates the read and write counts of the
traffic that traverses it, multiplying the counters by a decay rate $\gamma
  \in (0,1)$ at regular intervals $\Delta t$ so that recent traffic dominates.
These decayed metrics are aggregated bottom-up only when the SR data placement
protocol runs (\s{sr-dataPlacement}).

\subsection{Data Placement}
\label{sec:dec-placement}

Data placement is a continuous adaptation: as invocations arrive, each node
observes the demand that traverses it and adjusts its replicas accordingly,
reclaiming space when storage runs short through a hierarchical eviction
policy shared by both consistency models (\s{dec-eviction}).
The asymmetry of the two consistency models translates into two mechanisms: SR
placement is proactive, with a periodic protocol deciding leader and followers
from tree-aggregated demand (\s{sr-dataPlacement}); ER placement is reactive,
with new replicas emerging as a side effect of serving requests
(\s{er-dataPlacement}).

\subsubsection{Proactive Placement (SR)}
\label{sec:sr-dataPlacement}

\begin{figure*}[htbp]
  \centering

  \begin{subfigure}{0.48\textwidth}
    \centering
    \resizebox{0.6\columnwidth}{!}{%
      \begin{tikzpicture}[ host/.style={circle, draw, thick, minimum
              size=1.5cm, fill=gray!5}, rep/.style={draw, rounded corners=2pt,
              inner sep=3pt, font=\footnotesize\sffamily, fill=gray!15},
          demand/.style={draw, rounded corners=3pt, thick, align=center,
              font=\scriptsize, minimum width=1.7cm, minimum height=0.85cm,
              fill=white}, spring/.style={decorate, decoration={zigzag,
                  segment length=5pt, amplitude=2.5pt, pre length=5pt, post
                  length=7pt}, thick, -{Stealth[scale=1.1]}},
          buoy/.style={-{Stealth[scale=1.2]}, very thick, dashed,
              blue!55!black}, flbl/.style={font=\scriptsize, align=center} ]
        \node[host] (nj) {}; \node[rep] (rep) at (nj) {$c$};
        \node[font=\scriptsize\itshape, below=2pt of nj.south] {$j$};

        \node[demand] (w) at (4.1, 1.2) {writers\\ $\lambda^W$}; \node[demand]
        (r) at (4.1, -1.2) {readers\\ $\lambda^R$};

        \draw[spring, red!75!black] (rep.east) -- (w.west); \node[flbl,
          text=red!75!black] at (2.35, 1.45)
        {$F_{\text{elastic}}(\lambda^W\!\!,d)$}; \draw[spring,
          green!55!black] (rep.east) -- (r.west); \node[flbl,
          text=green!55!black] at (2.35, -1.45)
        {$F_{\text{elastic}}(\lambda^R\!\!,d)$};

        \draw[buoy] (nj.north) -- ++(0, 1.6); \node[flbl, text=blue!55!black,
          anchor=south] at (0, 2.45) {$F_{\text{buoyancy}}(j)$};
      \end{tikzpicture}%
    }
    \caption{Virtual forces acting on a replica}
    \label{subfig:virtual_forces}
  \end{subfigure}
  \hfill
  \begin{subfigure}{0.48\textwidth}
    \centering
    \resizebox{0.8\columnwidth}{!}{%
      \begin{tikzpicture}[ nodec/.style={circle, draw, thick, minimum
              size=1.15cm, fill=white}, faint/.style={circle, draw, thin,
              gray!45, minimum size=0.8cm, fill=gray!3}, rep/.style={draw,
              rounded corners=2pt, inner sep=1.5pt, font=\scriptsize\sffamily,
              fill=gray!15}, spring/.style={decorate, decoration={zigzag,
                  segment length=4pt, amplitude=2pt, pre length=4pt, post
                  length=6pt}, thick, -{Stealth[scale=1]}},
          tens/.style={-{Stealth[scale=1.2]}, line width=1.1pt,
              red!70!black}, lbl/.style={font=\scriptsize, align=center},
          elbl/.style={font=\tiny\itshape, gray} ]

        \node[nodec] (nj) at (0, 2.1) {}; \node[rep] at (nj) {$\bigstar$};
        \node[font=\scriptsize, anchor=west] at ([xshift=2pt]nj.east) {$j$};
        \node[nodec] (nk) at (-2.7, 0.3) {$k$}; \node[nodec] (ni) at ( 2.7,
        0.3) {$i$};

        \draw[tens] (0, 3.8) -- (nj.north); \node[lbl, text=red!70!black,
          anchor=south] at (0, 3.85) {$F_{\text{ext},j}^W$ (inherited)};

        \draw[spring, teal!70!black] (nj.south west) -- (nk.north east);
        \node[lbl, text=teal!70!black, anchor=east] at (-3.5, 0.3)
        {$F_{\text{elastic}}$\\ $(\lambda^W_k,\mathit{lat}_{jk})$};

        \draw[spring, red!70!black] (nj.south east) -- (ni.north west);
        \node[lbl, text=red!70!black, anchor=west] at (3.5, 0.3)
        {$F_{\text{elastic}}$\\ $(\lambda^W_i,\mathit{lat}_{ji})$};

        \node[lbl, fill=white, inner sep=1.5pt] at (0, 1) {move leader\\ to
          $k$? \eqref{eq:leader-delegation}};

        \draw[tens, dashed] (-2.7, 1.5) -- (nk.north); \node[lbl,
          text=red!70!black, anchor=south] at (-2.7, 1.6)
        {$F_{\text{ext},k}^W$ \eqref{eq:ancestor-tension}};

        \node[faint] (g1) at (-3.3, -0.9) {}; \node[faint] (g2) at (-2.1,
        -0.9) {}; \draw[thin, gray!45] (nk) -- (g1); \draw[thin, gray!45] (nk)
        -- (g2); \node[elbl, anchor=north] at (-2.7, -0.35) {recurse};
      \end{tikzpicture}%
    }
    \caption{Recursive tension evaluation for leader delegation}
    \label{subfig:tension_propagation}
  \end{subfigure}

  \caption{Proactive data placement model for SR collections. (a) A replica
    feels opposing virtual forces: elastic forces pull it toward the read and
    write demand it serves, while a buoyant force expels it from nearly-full
    nodes. (b) Recursive evaluation for leader ($\bigstar$) delegation: node
    $j$ weighs the elastic pull from a write-heavy child $k$ against the
    opposing sibling forces and the inherited tension $F_{\text{ext},j}^W$; on
    migration these fold into a single scalar $F_{\text{ext},k}^W$ passed to
    $k$.}
  \label{fig:sr_forces_model}
\end{figure*}
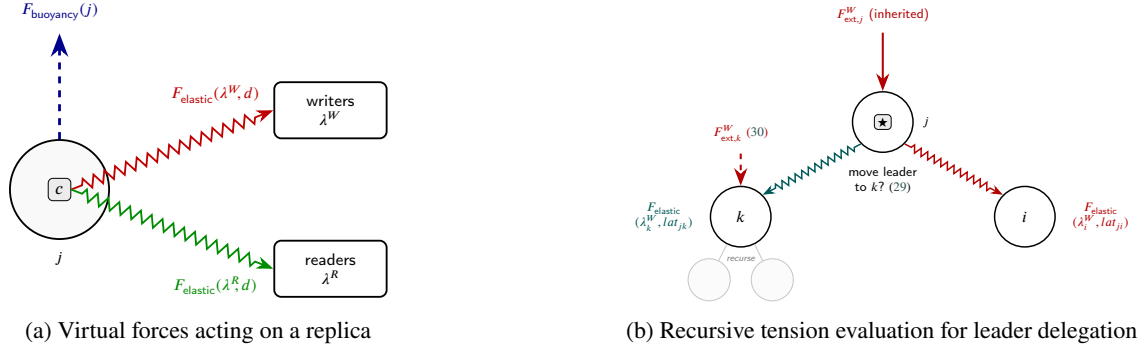

\fakeparagraph{Protocol}
A periodic protocol supplies, epoch by epoch, the aggregated metrics that the
SR placement conditions consume; ER collections take no part, being purely
reactive. Each epoch comprises three phases
(Algorithm~\ref{alg:placement-cycle}): a measurement phase (\emph{Idle}), a
bottom-up aggregation (\emph{Gossip}), and a top-down decision
(\emph{Decision}).

The two directions follow from how demand and information are distributed.
Clients issue requests at every level, each pulling the collections it
accesses toward itself; summarizing these pulls is inherently bottom-up, so a
node's view of demand grows more complete the closer it sits to the root.
Decisions therefore flow downward, each node resolving its subtree and
supplying every child the external context it cannot observe on its own.

\begin{algorithm}[t]
  \footnotesize
  \caption{Placement protocol cycle at node $j$.}
  \label{alg:placement-cycle}
  \begin{algorithmic}[1]
    \Statex \textbf{Idle phase:} \State accumulate per-collection read/write
    counters for duration $\Delta t$

    \Statex \Statex \textbf{Gossip phase (bottom-up):} \State wait for the
    summary $s_k$ from every child $k \in \mathrm{ch}(j)$ \State $s_j \gets$
    \Call{Aggregate}{local counters, $\{s_k\}_{k \in \mathrm{ch}(j)}$, storage
      util.} \If{$j \neq \mathrm{root}$} \State send $s_j$ to $\mathrm{par}(j)$
    \EndIf

    \Statex \Statex \textbf{Decision phase (top-down):} \ForAll{SR collections
      $c$ with $\mathsf{local}\in\rho_j(c)$} \ForAll{children $i \in
        \mathrm{ch}(j)$} \State \Call{EvalLeaderDelegation}{$c, i$}
    \Comment{Eq.~\eqref{eq:leader-delegation}} \State
    \Call{EvalFollowerReplication}{$c, i$}
    \Comment{Eq.~\eqref{eq:follower-replication}} \If{a replica is
      provisioned on a full node} \State \Call{Evict}{$\cdot$}
    \Comment{\s{dec-eviction}} \EndIf \EndFor \EndFor \State send outcomes
    to each child $i \in \mathrm{ch}(j)$
  \end{algorithmic}
\end{algorithm}

\fakeparagraph{Physical model}
We reason about a single SR collection $c$ at a time: every demand aggregate
and force introduced below is relative to $c$, a dependence we leave implicit
to lighten the notation. SR placement must jointly decide
\begin{inparaenum}[(1)]
  \item where the single leader sits, and
  \item how many followers to maintain.
\end{inparaenum}
We model both as a balance of \emph{virtual forces} (\fig{sr_forces_model}).

Drawing on the spring analogy of Pietzuch et al.~\cite{pietzuch:networkAware},
the attraction between a replica and the demand it serves is an \emph{elastic}
force: the product of the request rate $\lambda$ and the latency $d$ between
them \eqref{eq:forces}, where $d$ is the network path latency $\mathit{lat}$.
Since writes are served only by the leader whereas any replica serves reads,
we decompose demand accordingly: the write rate $\lambda^W$ pulls only on the
leader, whereas the read rate $\lambda^R$ pulls on whichever replica is
nearest, and thus governs where followers appear.

Storage pressure is different in kind: it carries no notion of distance and
depends only on node utilization. We model it as a \emph{buoyant} force that,
like a fluid expelling an immersed body, grows with the collection size and
with how full the host is \eqref{eq:forces}; the constant $\mu$ aligns the
storage and latency scales so the two forces are comparable.

  {\footnotesize
    \begin{align}
      F_{\text{elastic}}(\lambda, d) & = \lambda \cdot d,
      \qquad
      F_{\text{buoyancy}}(j) = \frac{\mu}{\mathit{stor}^{\text{free}}_j}
      \cdot \mathit{size}_c
      \label{eq:forces}
    \end{align}
  }

\fakeparagraph{Force balance}
A replica relocates only when the net force is favorable, so the protocol
converges to a \emph{force equilibrium}: the leader comes to rest at the
barycenter of write traffic, while a follower materializes wherever read
demand overcomes the local storage resistance.
This is the same proximity-versus-storage tradeoff the BLP optimizes globally
from a complete snapshot; here each node applies it locally and incrementally,
without ever materializing a global objective.

\fakeparagraph{Demand aggregates}
The placement conditions introduced below rely on the traffic aggregates
supplied by the Gossip phase. For a node $k$, we define the rate-weighted
write and read demand that $c$ draws from the subtree of $k$ as:

{\footnotesize
  \begin{align}
    \lambda^W_k = \!\!\sum_{i \in \mathrm{sub}(k)} \sum_{\substack{f \in \Fw \\ c \in
                                                       \setC_f}} \!\!\lambda_{f,i},
    \qquad
    \lambda^R_k = \!\!\sum_{i \in \mathrm{sub}(k)} \sum_{\substack{f \in \Fr \\ c \in
                                                       \setC_f}} \!\!\lambda_{f,i},
    \label{eq:subtree-demand}
  \end{align}}

\fakeparagraph{Leader delegation}
Moving the leader from node $j$ to a child node $k$ compares two elastic
forces against each other (\fig{sr_forces_model}(b)): the pull from $k$'s
write-heavy subtree favors the move, while the pull from every writer outside
that subtree ($k$'s siblings and all demand above $j$) opposes it. We call
this latter aggregated force \emph{tension} at $j$, denoted
$F_{\text{ext},j}^W$. The leader migrates to $k$ when
condition~\eqref{eq:leader-delegation} holds: on the left, the net inward
pull, i.e. the elastic force from $k$ minus the opposing sibling forces minus
the inherited tension; on the right, the buoyancy cost of placing the replica
one tier deeper, the extra hop that sibling writes must now travel, and a
hysteresis threshold $\tau_L$ preventing oscillation between adjacent
placements.

  {\footnotesize
    \begin{align}
       & F_{\text{elastic}}(\lambda^W_k, \mathit{lat}_{jk})
      - \!\!\sum_{\substack{i \in \mathrm{ch}(j) \\ i \neq k}}\!\!
      F_{\text{elastic}}(\lambda^W_i, \mathit{lat}_{ji}) - F_{\text{ext},j}^W
      \nonumber                                             \\
       & \qquad > F_{\text{buoyancy}}(k)
      + \!\!\sum_{\substack{i \in \mathrm{ch}(j) \\ i \neq k}}\!\!
      \lambda^W_i \cdot \mathit{lat}_{jk} + \tau_L
      \label{eq:leader-delegation}
    \end{align}}

If the leader moves, the same test repeats at node $k$ against its own
children. For the condition to retain the same form at every depth, node $k$
needs a single tension scalar summarizing all write demand outside its
subtree. Node $j$ constructs it by folding its sibling elastic forces and
their hop penalties into the inherited $F_{\text{ext},j}^W$, yielding the
$F_{\text{ext},k}^W$ \eqref{eq:ancestor-tension} that it hands down to $k$
(\fig{sr_forces_model}(b)). The recursion therefore propagates one tension
value downward, and every delegation decision has identical structure
regardless of depth.

  {\footnotesize
    \begin{align}
      F_{\text{ext},k}^W \gets
      \!\!\sum_{\substack{i \in \mathrm{ch}(j) \\ i \neq k}}\!\!
      F_{\text{elastic}}(\lambda^W_i, \mathit{lat}_{ji})
      + \!\!\sum_{\substack{i \in \mathrm{ch}(j) \\ i \neq k}}\!\!
      \lambda^W_i \cdot \mathit{lat}_{jk}
      + F_{\text{ext},j}^W
      \label{eq:ancestor-tension}
    \end{align}}

\fakeparagraph{Follower replication}
Followers serve reads near the edge. The decision is local to the candidate
subtree, with no sibling term: node $j$ provisions a replica on child node $k$
when the read demand from $k$'s subtree overcomes the buoyancy there
\eqref{eq:follower-replication}, with $\tau_R$ a threshold.

  {\footnotesize
    \begin{align}
      F_{\text{elastic}}(\lambda^R_k, \mathit{lat}_{jk}) > F_{\text{buoyancy}}(k)
      + \tau_R
      \label{eq:follower-replication}
    \end{align}}

\subsubsection{Reactive Placement (ER)}
\label{sec:er-dataPlacement}

ER placement reduces to a single \emph{always-copy} policy. When a function
executes on a node $j$, missing one of its ER collections
that collection is fetched from the nearest reachable replica, located by
following $\rho_j(c)$ through the reachability table (\s{dec-state}); the
serving replica then asynchronously spawns a fresh copy one hop closer to the
consumer. A collection under sustained demand therefore migrates hop by hop
toward the edge nodes that access it. Hierarchical eviction (\s{dec-eviction})
provides the opposing pull: copies that are large and rarely accessed are
reclaimed first, drifting back toward higher tiers.

\subsubsection{Hierarchical Eviction}
\label{sec:dec-eviction}

Both consistency models replicate aggressively toward the edge and rely on a
shared eviction mechanism to stay within node capacity. A node that must host
a replica it does not already hold, without enough room for it, repeatedly
evicts the replica of highest priority $\mathit{size}_r \cdot (t -
  t_j^{\text{last}}(r))$, with $t$ the current time, until the incoming replica
fits: large, stale replicas leave first, and leaders are exempt
(\s{dec-state}). Each evicted replica is demoted to the parent, so storage
pressure propagates upward one hop at a time until it is absorbed by a node
with spare capacity or by the cloud, whose unlimited storage guarantees
termination. A node that receives a replica it already holds discards the
incoming copy as redundant. Eviction also keeps the local state of
\s{dec-state} consistent: when a replica leaves a node and no copy remains in
its subtree, the node has its parent drop the stale reachability entry.

\subsection{Function Scheduling}
\label{sec:dec-scheduling}

Placement decides where replicas live; scheduling routes each invocation to
one. A node reads its reachability table (\s{dec-state}) and either serves the
request or forwards it one hop, so each invocation discovers its target by
walking the hierarchy (Algorithm~\ref{alg:scheduling}).

Consistency dictates that target. An SR access is the stricter of the two and
takes priority: a write must reach the \emph{leader}, a read any replica, so
the node follows the single known direction to that target ($\ell_j(c)$ or
$\rho_j(c)$) and climbs to $\mathrm{par}(j)$ only when it lies upstream. An
ER-only invocation instead seeks maximal co-location: it executes wherever any
required collection is already local, and otherwise descends toward the child
whose subtree covers the most of them, falling back upstream when none is
reachable below. Any collections still missing at the execution node are then
fetched from the nearest known replica (\s{er-dataPlacement}).

\begin{algorithm}[t]
  \footnotesize
  \caption{Function scheduling at node $j$.}
  \label{alg:scheduling}
  \begin{algorithmic}[1]
    \Require Function $f$ arriving at node $j$ \If{$\CfSR \neq \emptyset$}
    \State $t \gets \{\ell_j(c)\}$ \textbf{if} $f \in \Fw$ \textbf{else}
    $\rho_j(c)$

    \If{$\mathsf{local} \in t$} \State execute $f$ on $j$; \Return \ElsIf{$t
        \cap \mathrm{ch}(j) \neq \emptyset$} \State forward $f$ to a child in
    $t$; \Return \Else \State forward $f$ to $\mathrm{par}(j)$ \EndIf
    \Else \Comment{ER-only access} \If{$\exists\, c \in \setC_f :
        \mathsf{local} \in \rho_j(c)$} \State execute $f$ on $j$; \Return
    \ElsIf{$\exists\, c \in \setC_f,\ \exists\, k \in \mathrm{ch}(j) : k
        \in \rho_j(c)$} \State $k^\ast \gets \displaystyle\arg\max_{k \in
        \mathrm{ch}(j)} \bigl|\{c \in \setC_f : k \in \rho_j(c)\}\bigr|$
    \State forward $f$ to $k^\ast$; \Return \Else \State forward $f$ to
    $\mathrm{par}(j)$ \EndIf \EndIf
  \end{algorithmic}
\end{algorithm} %
\section{Evaluation}
\label{sec:eval}

The evaluation asks whether BLP, GV, and AV offer a viable path from
theoretical optimality to runtime deployment in edge-to-cloud settings. To
this end, we structure our analysis around four main Research Questions (RQs).
We first establish the practical limits of exact optimization, asking how far
the BLP can scale before becoming intractable and how GV compares along the
same dimensions (RQ1); only within that solvable regime can we then measure
how much solution quality GV and AV sacrifice relative to the optimum (RQ2).
Both rest on the BLP objective, which scores a placement decision but not the
operational reality of enforcing it. We therefore complement the analytic
comparison with the latency that clients actually observe, under a static
workload (RQ3) and under shifting access patterns (RQ4).

\fakeparagraph{\textbf{RQ1}}
To what extent do system dimensions (number of nodes, collections, functions)
and consistency requirements affect the scalability of the BLP in terms of
solve time and memory footprint? How does GV scale along the same system
dimensions?

\fakeparagraph{\textbf{RQ2}}
How closely do GV and AV approximate the quality of exact optimization of the
BLP in terms of total latency and total storage? How do they compare to
representative deployment baselines?

\fakeparagraph{\textbf{RQ3}}
Under a static workload, what client-observed latency do BLP, GV, AV, achieve
in a realistic edge-to-cloud deployment?

\fakeparagraph{\textbf{RQ4}}
When access patterns shift over time, how does the client-observed latency of
the same policies respond?

\subsection{Experimental Setup}
\label{sec:eval-setup}

Our evaluation has two complementary needs. RQ1 and RQ2 assess scalability and
solution quality on static problem instances. BLP and GV produce placement
decisions directly from the problem snapshot, and their cost is read from the
optimizer's own analytical objective. AV, being a runtime heuristic, requires
a simulation to converge to a placement; its cost is then computed
analytically using the same objective function, so that all three approaches
are compared on the same metric (\s{eval-scalability}).
RQ3 and RQ4 shift the perspective from decision quality to operational
behavior: all policies, including BLP and GV, are embedded in a discrete-event
simulator that measures the latency clients actually experience, under a
static workload for RQ3 and under a shifting one for RQ4.

\fakeparagraph{Baselines and Competitors}
Our system model differs fundamentally from prior work (\s{related}),
therefore a direct algorithmic comparison with existing solutions is
unfeasible. Instead, in line with similar evaluation strategies in the
literature~\cite{nardelli:functionOffloading,puliafito:stateAllocation}, we
evaluate our policies against two representative deployment baselines.
A \emph{Cloud-only} (\emph{CO}) policy places a single replica of every
collection at the cloud, schedules every invocation there, and elects the
cloud as leader for all SR collections. It represents the canonical FaaS
deployment, with compute and data co-located within the data center.
A \emph{Cloud-data} (\emph{CD}) policy retains CO data placement but executes
ER-only invocations at the edge (walking upward in the tree to the first
ancestor with sufficient capacity if needed); SR invocations are forced to the
cloud by co-location, and every ER access from a non-cloud node triggers a
remote fetch from the cloud. CD captures the naive extension of FaaS to the
edge, where compute follows the user, but data stays centralized.

\fakeparagraph{Centralized Testbed}
The centralized testbed is implemented in C++. It comprises the BLP model,
implemented using the Concert Technology of IBM ILOG
CPLEX\footnote{\url{https://www.ibm.com/products/ilog-cplex-optimization-studio}},
the GV heuristic, and the baselines. Given a generated problem instance, this
testbed produces a placement and scheduling decision.
A 600\,s timeout is enforced on the BLP solver, and virtual memory is capped
at 50\,GB.

\fakeparagraph{Simulator Testbed}
The BLP and GV produce a placement from a single snapshot, but do not capture
how that placement performs over time as invocations arrive, clients move, and
the solver re-plans. To evaluate operational behavior without the cost of a
full deployment, we embed all policies in a discrete-event simulator in
Go\footnote{\url{https://github.com/fschuetz04/simgo}}.
A set of workload generators continuously emit function invocations according
to configurable arrival patterns. Each invocation specifies a source node, the
set of accessed collections and their consistency requirements, and a
reference execution time. The active policy processes each invocation by
selecting an execution node and resolving data access; the simulator then
records the resulting latency observed by clients.

Centralized policies run the solver in a continuous loop. At each iteration it
reads the invocations falling within a sliding window of the most recent $W$
seconds, estimates the rates $\lambda_{f,i}$ by dividing the observed counts
by $W$. It then computes a new placement plan, and advances the simulation
clock by the wall-clock solve time plus the cost of two tree traversals: a
gather that collects the system state and a scatter that distributes the plan,
each charged as the one-way latency from root to the deepest leaf. It then
immediately begins the next iteration, on a window that now reflects the
invocations arrived during the previous solve cycle. We deliberately charge
only this control-plane cost: relocating replicas to their target nodes would
also incur a data-plane cost, which the simulator omits, applying the new
placement instantaneously.
Invocation latency is computed analytically from the active plan: round-trip
delay to the execution node, function execution time, and, for ER collections
not co-located with the function, the retrieval cost from the provider
replica.

Under AV no global plan exists: each node reacts to invocations independently,
routing them hop-by-hop through the hierarchy (\s{dec-scheduling}) while data
placement evolves continuously (\s{dec-placement}). Latency is therefore
measured end-to-end as each invocation traverses the simulated network,
accumulating discovery, retrieval, execution, and return-path delays. Unlike
the centralized policies, AV pays the data-plane cost explicitly, migrating
replicas hop by hop as it serves requests. The centralized latencies reported
below are therefore optimistic, a bias in their favor that only strengthens
the comparison wherever AV matches or outperforms them.

\fakeparagraph{Machine Specifications}
All experiments were conducted on a server equipped with a 16-core AMD Ryzen 9
9950X processor (32 threads) and 64\,GB of DDR5 RAM, running Fedora Server 42.

\fakeparagraph{Scenario Generation}
All four research questions share a common scenario generation procedure. RQ1
and RQ2 vary the problem dimensions (nodes, functions, collections); RQ3 and
RQ4 fix them at their defaults and vary client mobility.
We ground the scenario in empirical benchmarks from the literature rather than
choosing parameters arbitrarily: network and topology characteristics follow
the edge-cloud taxonomy of Belcastro et
al.~\cite{belcastro:edgeCloudContinuum}, while function execution times and
data sizes are derived from the EdgeFaaSBench
suite~\cite{rajput:EdgeFaaSBench:ieeeEdge:2022}, a benchmark for stateful
serverless functions on constrained edge hardware.
We instantiate a three-tier topology (cloud, fog, edge) as a balanced $b$-ary
tree of depth~3. The default $b = 10$ yields one cloud root, 10 fog nodes, and
100 edge nodes (111 total), matching the device-density ratios of Belcastro et
al.~\cite{belcastro:edgeCloudContinuum}; when $|\setN|$ is varied (RQ1), $b$
is recomputed from $|\setN| = 1 + b + b^2$ to preserve the three tiers.
Network parameters follow the per-tier ranges of the same
work~\cite{belcastro:edgeCloudContinuum} (our cloud/fog/edge map to their
cloud/near-edge/far-edge): bandwidth MBps--Gbps (edge), Gbps (fog), Gbps--Tbps
(cloud); latency 2--5, 10--20, and ${\geq}\,20$\,ms. Taking the geometric mean
of each range as its representative value gives per-tier bandwidths of
$\approx$\,90\,Mbps (edge) and $\approx$\,1\,Gbps (fog), and latencies of
$\approx$\,3.2\,ms (edge) and $\approx$\,14.1\,ms (fog). Inter-tier links take
the minimum bandwidth and the summed latency of their endpoints: 0.011\,MB/ms
and 17.3\,ms (edge--fog), 0.125\,MB/ms and 34.1\,ms (fog--cloud).
Storage capacity is expressed as a fraction of the total collection size,
hence independent of absolute scale: cloud nodes are unlimited, fog nodes host
50\% and edge nodes 30\% of the workload, calibrated against the access
pattern below.
Reference execution times are drawn from an exponential distribution with mean
279\,ms. The suite reports end-to-end latencies for stateful serverless
functions on constrained IoT hardware (Raspberry~Pi~4, Jetson~Nano); we retain
the four benchmarks compatible with persistent-state access (image processing,
speech-to-text, image classification, object detection, all CPU) and their
warm-start measurements, which average $\approx$\,837\,ms across benchmarks
and devices.
Since these target low-power ARM cores (Cortex-A72) whereas our fog tier
models server-grade processors (Intel Xeon~E5), $\sim3\times$ faster in
single-thread
benchmarks~\footnote{\href{https://www.cpubenchmark.net/compare/4078vs1217/ARM-Cortex-A72-vs-Intel-Xeon-E5-2643}{CPU
    Benchmark: ARM Cortex-A72 vs Intel Xeon E5-2643}}, we apply this factor to the
empirical mean to obtain the 279\,ms baseline.
Processing speed is a per-tier factor normalized to the fog baseline
($1\times$): cloud $1.1\times$, edge $0.9\times$. This ${\pm}10\%$ gradient
reflects the comparable single-thread performance of server-grade processors
across tiers, and keeps the placement driven by topology and data locality
rather than compute asymmetry.

\begin{figure}
  \centering
  \includegraphics[width=\columnwidth]{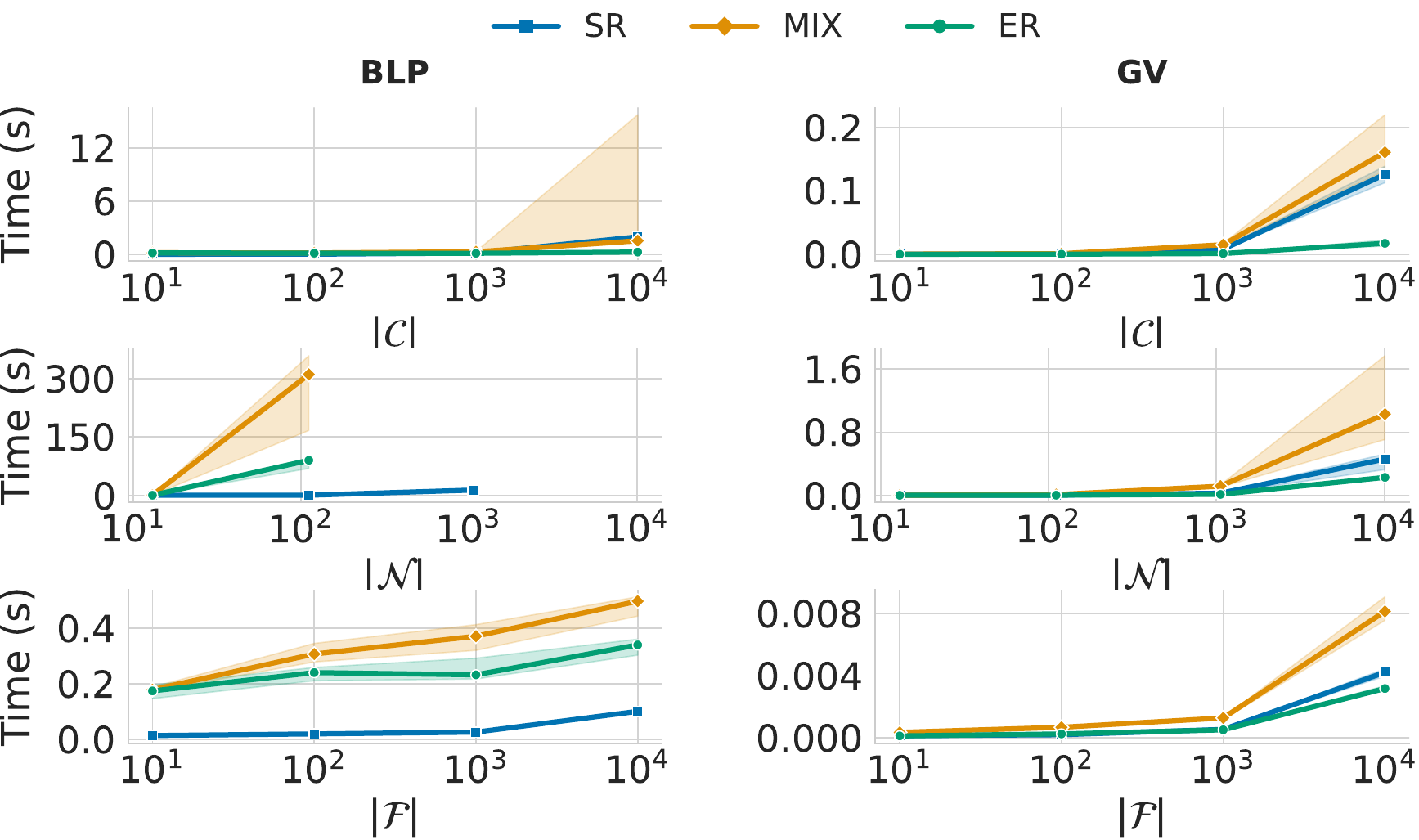}
  \caption{Solve time ($\alpha = 1$)}
  \label{fig:scalability-time}
\end{figure}

\fakeparagraph{Workload Generation}
Since the state accessed by these benchmarks consists of image frames or audio
snippets, we draw collection sizes from an exponential distribution with mean
1\,MB, small enough that the edge--fog bandwidth (90\,Mbps) is not an
artificial bottleneck.
Each function accesses collections sampled uniformly from the pool: 1--2 ER
collections (a maximum local working set of $\approx$\,2\,MB), or exactly one
for functions accessing SR. To model mutations, 80\% of functions are
read-only and 20\% writing.
This justifies the 30\% edge storage capacity set earlier: with 10 collections
($\approx$\,10\,MB total, per \s{eval-scalability}), 30\% equals 3\,MB,
exactly the maximum working set of one function execution, so an edge node
avoids immediate cache thrashing.
Invocation traffic comes from 100 virtual clients, each selecting a function
by a Zipf distribution (skew~$1.2$), a source node tier-weighted 70\% edge /
30\% fog, and a rate drawn from an exponential distribution with
mean~$0.1$\,req/s. Rates of clients sharing a (function, node) pair accumulate
into one entry of $\lambda_{f,i}$, and each active pair becomes a simulator
client with that aggregated rate.
All collections are initially placed at the cloud, reflecting a cold start
from a centralized deployment.

\subsection{Scalability of Exact and Greedy Optimization}
\label{sec:eval-scalability}

To answer RQ1, we evaluate how BLP and GV scale as we independently vary the
number of nodes $|\setN|$, functions $|\setF|$, and collections $|\setC|$. For
each dimension we scale one parameter while holding the other two fixed at the
default values ($|\setN| = 13$, $|\setF| = 10$, $|\setC| = 10$). The node
sweep uses $|\setN| \in \{13, 111, 1057, 10101\}$, corresponding to branching
factors $b \in \{3, 10, 32, 100\}$ with a fixed tree depth of~3. A fixed depth
ensures that the three-tier structure (and its associated network and speed
parameters) remains invariant across the sweep, isolating the effect of scale
from changes in topology shape. Functions and collections are swept over
$\{10, 100, 1000, 10000\}$.
AV is excluded from these scalability plots because it is evaluated via
simulation.
We further vary the consistency mix of the workload: \emph{ER} (only
eventually-replicated collections), \emph{SR} (only strongly-replicated), and
\emph{MIX} (50/50).
The entire sweep is repeated for two settings of the latency--storage
trade-off, $\alpha \in \{1, 0.01\}$. Since both terms of the objective
are normalized, $\alpha = 1$ weighs storage and latency equally; we do not go
beyond this value, as we are not interested in regimes where storage dominates
latency, but only in those where it counts as much as, or less than, latency.
Each resulting configuration is solved with $10$ different random seeds; we
report the median and interquartile range (IQR) of total wall-clock execution
time in \fig{scalability-time}; missing markers either exceeded the 50\,GB
memory cap or the 600\,s timeout.

\fakeparagraph{Results}
We report results for $\alpha = 1$. Preliminary tests confirmed that varying
$\alpha$ does not affect solve time or memory, since it only rescales one term
of the objective without altering the constraint structure or the size of the
search space.
Across all three dimensions, MIX consistently requires the longest solve time
for both BLP and GV. This is expected, as the interaction between SR placement
constraints and ER replication freedom enlarges the effective search space.
Functions are the least demanding dimension for both BLP and GV. Each
additional function introduces scheduling variables, but scheduling decisions
are independent of one another: where one function executes does not constrain
where another can run.
BLP solves all configurations within 0.6\,s (MIX), 0.4\,s (ER), and 0.1\,s
(SR). GV is consistently faster: 0.20\,s (MIX), 0.08\,s (SR), and below
0.01\,s (ER).
Collections, by contrast, are harder to scale. Each additional collection adds
$|\setN|$ placement variables ($x_{c,j}$) together with storage and
consistency constraints that couple it to every other collection on the same
node. BLP exhibits high variance under MIX, with a median of $\approx$\,2\,s
but IQR peaks near 16\,s at $10^4$ collections. This variability arises
because the random assignment of SR and ER labels to collections produces
instances of uneven difficulty: seeds where the SR/ER split creates tighter
interactions between placement and routing take disproportionately longer, as
the branch-and-bound search explores more nodes before converging. SR
instances also reach $\approx$\,2\,s but with narrow IQR, confirming that the
uniform constraint structure of a single consistency model leads to more
predictable solver behavior. ER remains below 1\,s, as the absence of leader
constraints simplifies the feasible region.
GV follows the same ordering but at much lower absolute values: MIX reaches
0.26\,s with modest variability, SR 0.18\,s, and ER $\approx$\,0.01\,s.
Nodes are the most demanding dimension, as the number of decision variables
grows quadratically: the scheduling variable $w_{f,i,j}$ and the network
parameter matrices (latency, bandwidth) scale with $|\setN|^2$. BLP fails to
complete at $|\setN| = 10101$ for any consistency mix. SR survives to $|\setN|
  = 1057$ thanks to its reduced variable count, which keeps memory within the
cap. However, at that scale the remaining free variables still produce a
search space that exceeds the 600\,s timeout. ER and MIX fail earlier
($|\setN| = 111$) because without co-location pruning, the full variable set
exhausts the 50\,GB memory cap before the solver can even begin the search.
GV scales to $|\setN| = 10101$ for all consistency mixes. MIX is the most
expensive (120--240\,s, median 160\,s), followed by SR ($\approx$\,60\,s). ER
is remarkably cheap ($\approx$\,0.16\,s), because without leader constraints
the greedy evaluation of each candidate replica reduces to a simple
fetch-distance comparison.
For the memory footprint, we omit a dedicated figure, as its memory trends
mirror those of solve time. BLP memory is bounded by the CPLEX constraint
matrix: when scaling nodes, ER and MIX exhaust the 50\,GB cap at $|\setN| =
  111$, while SR survives to $|\setN| = 1057$ ($\approx$\,3\,GB) before reaching
the memory limit. When scaling collections, BLP reaches 2.8\,GB (SR), 2.0\,GB
(MIX), and 0.4\,GB (ER) at $10^4$. When scaling functions, all mixes stay
below 0.5\,GB.
GV uses two to three orders of magnitude less memory. Its footprint stays
below 0.014\,GB when scaling functions or collections, and is dominated by the
$|\setN|^2$ hop-count matrix when scaling nodes ($\approx$\,4.5\,GB at
$|\setN| = 10^4$).

\begin{figure*}
  \centering
  \includegraphics[width=\textwidth]{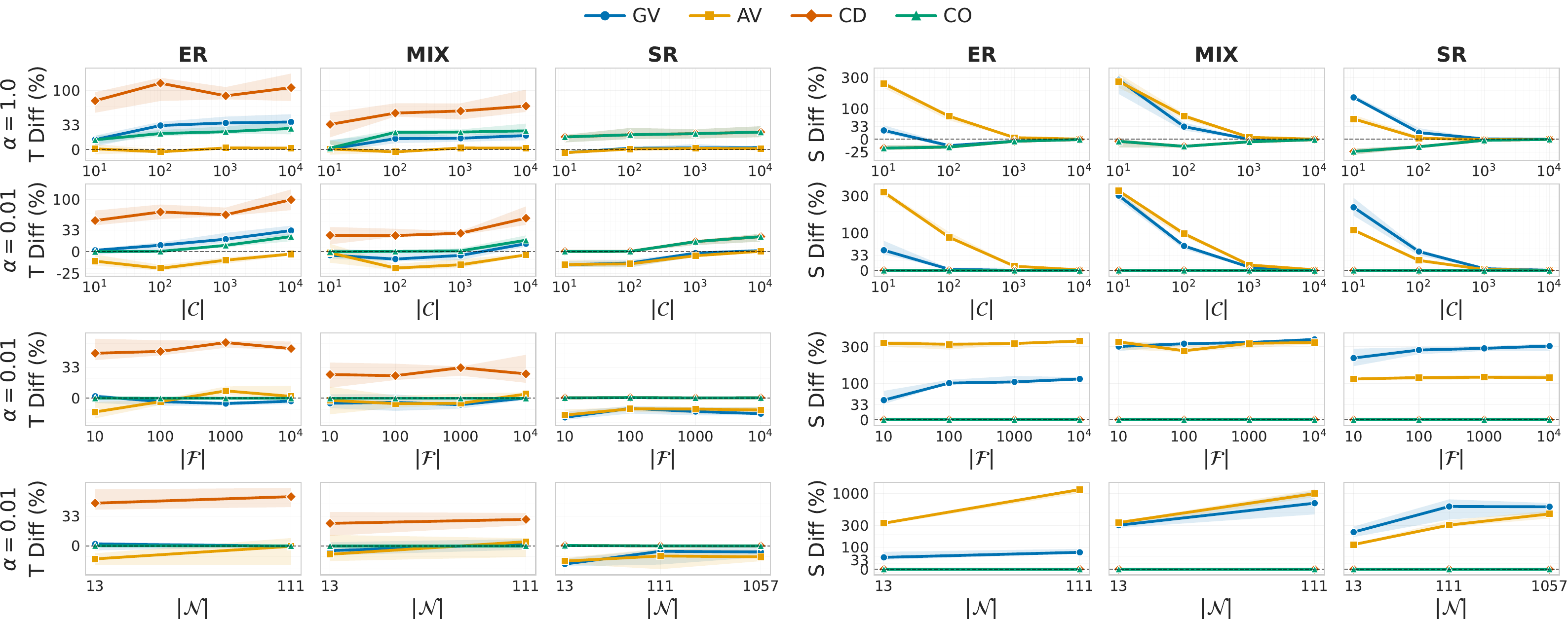}
  \caption{$T$ and $S$ difference relative to BLP optimum.}
  \label{fig:comparison}
\end{figure*}

\subsection{Solution Approximation}
\label{sec:eval-quality}

To answer RQ2, we reuse the same problem instances as RQ1, but instead of
their scaling behavior we now examine the quality of the placement each policy
produces. We evaluate GV and AV against the BLP optimum, complementing the
comparison with the aforementioned CO and CD baselines. Specifically, we
measure the rate-weighted total latency $T$ and the total storage cost $S$,
reporting the percentage difference of each heuristic and baseline relative to
the BLP optimum on the same instance. Each data point reports the median over
10 seeds; error bars span the IQR.

\fakeparagraph{Results}
In the collection sweep (\fig{comparison}), collections compete for finite
node storage, making the effect of $\alpha$ clearly visible.
Across all scenarios, the storage overhead of GV and AV relative to BLP is
large at small $|\setC|$ and converges toward zero as $|\setC|$ grows. The
additional replicas placed by these policies depend on the number of
\emph{accessed} collections, which is bounded by the function access pattern,
not by $|\setC|$. Unaccessed collections are treated identically by all
policies and cancel out in the difference, so the constant absolute surplus
shrinks as a percentage of the linearly growing baseline.
When latency and storage are weighted equally ($\alpha = 1$), replication is
expensive, and the optimal strategy is conservative.
Under ER, GV performs worse than CO in latency (T~diff up to $+18\%$ at
$|\setC| = 10$). We verified that the initial access-pressure placement
scatters single replicas across edge nodes; when a function accesses
collections in different subtrees, the scheduler selects the cloud as
execution point, triggering remote fetches through the narrow edge--fog links.
CO avoids this by co-locating all data and computation at the cloud from the
start. This is a structural limitation of GV when performing the greedy
expansion: the benefit of replicating a single collection is insufficient to
offset the total cost of fetching the remaining scattered dependencies.
AV achieves lower latency than both (T~diff below 10\%), because its
always-copy mechanism places replicas reactively regardless of storage cost.
This comes at a high storage premium ($\approx 250\%$ at $|\setC| = 10$),
which diminishes as $|\setC|$ grows for the reason discussed above.
CD is consistently the worst performer in latency (T~diff up to 95\%),
confirming that pushing compute to the edge without co-locating data is
counterproductive: every ER access triggers a remote fetch from the cloud.
Under SR, the picture changes. GV achieves near-optimal latency (T~diff within
$\pm 5\%$), because the co-location constraint fixes the execution node and
the greedy expansion effectively places follower replicas near readers. AV
matches this performance with comparable or lower storage. CO and CD are
forced to the cloud by co-location, incurring $\approx 15\text{--}18\%$ higher
latency.
Under MIX, the SR co-location constraint anchors the execution node, reducing
the ER component to a fetch-optimization problem where the greedy expansion is
effective. GV achieves near-optimal latency at small $|\setC|$ (median at
$0\%$, IQR up to $10\%$), with performance stabilizing around $+10\%$ at
$|\setC| = 10^4$. AV tracks a few percentage points above GV in latency, with
a similar storage trajectory.

As $\alpha$ decreases from 1 to 0.01, storage becomes cheaper in the
objective, and all policies replicate more aggressively.
Under ER, GV partially closes the gap with BLP as $\alpha$ decreases: at
$\alpha = 0.01$, the lower storage cost allows the greedy expansion to add
replicas that were previously rejected, reducing the scattering penalty. To
maintain its low computational footprint, GV evaluates candidates strictly one
at a time during its greedy expansion. Because of this, replicating a single
collection may yield no immediate benefit if the execution node remains fixed,
whereas jointly moving a cluster of co-accessed collections would unlock
further latency savings. Consequently, this incremental approach leaves GV
slightly above the BLP optimum.
Under SR and MIX, GV and AV both achieve latency within a few percentage
points of BLP at $\alpha = 0.01$.
The baselines degrade as $\alpha$ shrinks, because BLP now invests in replicas
that CO and CD cannot match. At $|\setC| = 10^4$, CO incurs $15\text{--}25\%$
higher latency across all mixes, and CD reaches $+99\%$ for ER, confirming
that centralized data with edge compute becomes increasingly suboptimal as the
placement budget relaxes.

Finally, the function and node sweeps are reported for $\alpha = 0.01$ in
\fig{comparison}; different values of $\alpha$ produce qualitatively similar
trends with smaller differences across policies. The node sweep is incomplete
beyond $|\setN| = 1057$ due to BLP memory limits. One notable exception in the
function sweep is visible under ER: as $|\setF|$ grows beyond 100, GV latency
decreases while storage increases, because each collection is accessed by a
larger aggregate rate, which pushes the greedy benefit above the $\alpha \cdot
  \text{size}_c$ threshold and triggers replication.

\begin{figure}
  \centering
  \includegraphics[width=\columnwidth]{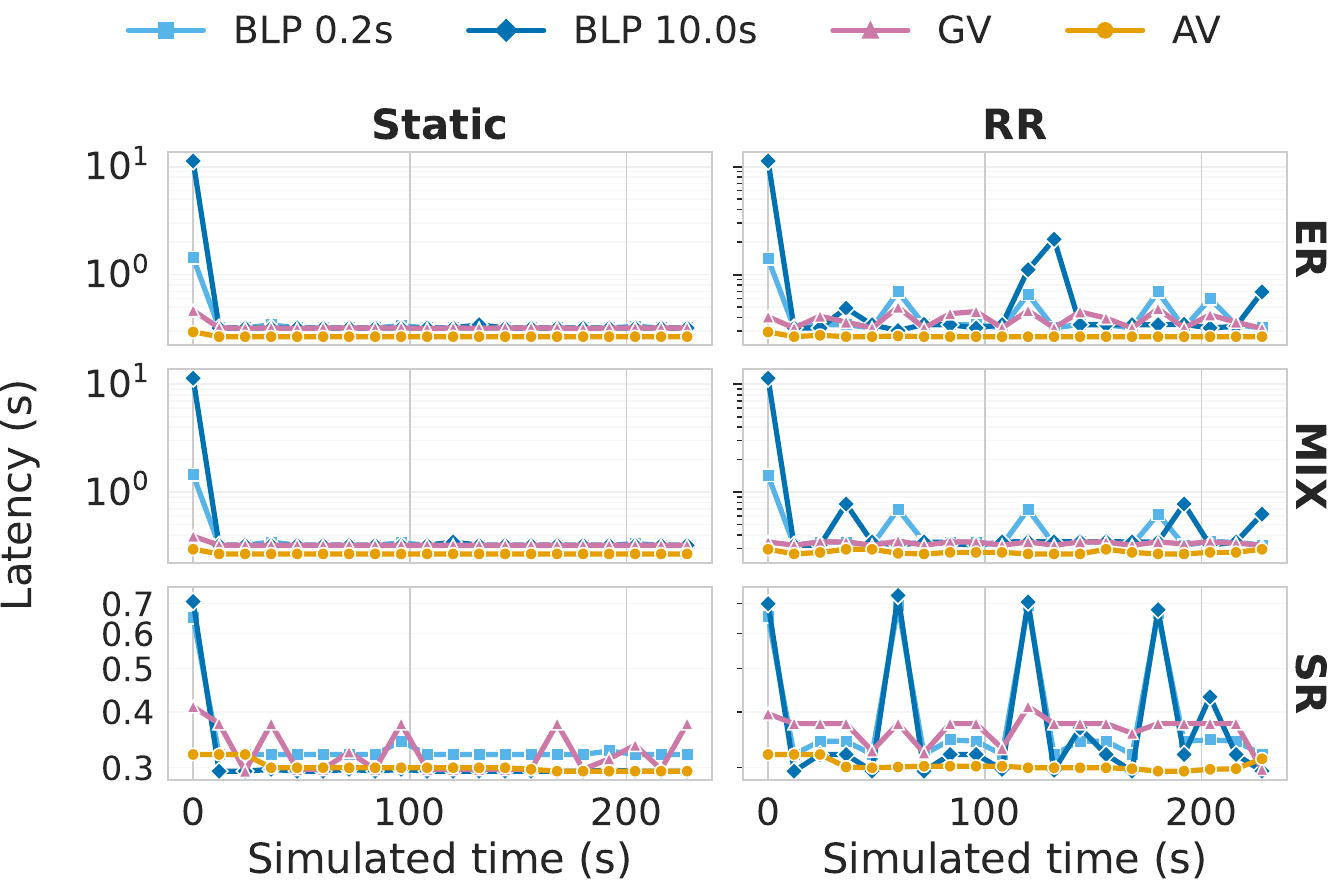}
  \caption{Client-observed latency for static workload (left) and mobile workload (right).}
  \label{fig:latency-combined}
\end{figure}

\subsection{Steady-state Behavior}
\label{sec:eval-decentralized}

To answer RQ3, we deploy all policies in the simulator under a static workload
with $|\setN| = 111$ ($b = 10$), $|\setF| = 100$, $|\setC| = 100$, and $\alpha
  = 0.01$. Clients invoke a given function at a constant rate from a fixed node
throughout the simulation.

Centralized policies use the sliding window described in \s{eval-setup} with
$W = 32$\,s. AV uses an identical 32\,s idle duration between protocol epochs.
The value is derived from the client arrival process: with the workload
configuration described above, a 32\,s observation window guarantees that at
least 70\% of all clients have emitted at least one invocation.
The BLP solver is evaluated under per-iteration timeouts of 0.2 and 10.0\,s;
in each case the best incumbent solution found within the budget is adopted,
regardless of its optimality gap.
Each simulation runs for 240\,s of simulated time.

\fakeparagraph{Results}
\fig{latency-combined} shows the median per-bin latency over time, with bins
of 12\,s.
All centralized policies exhibit an initial cold-start transient: the first
solver iteration operates on an empty or near-empty sliding window, producing
a suboptimal placement that penalizes the first few bins. Once the window
fills, the placement stabilizes and latency settles to its steady-state level.
AV does not suffer this transient. Its reactive placement begins adapting from
the first invocation, converging within the first epoch.
At steady state, BLP with 0.2\,s and 10\,s timeouts produce nearly
indistinguishable latency across all consistency mixes, suggesting that the
solver finds a good incumbent within the shorter budget. The main difference
is the cold-start transient: BLP~10s exhibits a higher initial spike because
the longer solve time delays the first plan, leaving more invocations
unoptimized during the first iteration.
Under ER and MIX, AV achieves lower median latency than all centralized
policies, because always-copy progressively co-locates the accessed
collections with the requesting edge nodes, eliminating remote fetches. Under
SR, all policies converge to comparable latency, as the co-location constraint
limits the placement space.
GV matches BLP under ER and MIX but exhibits periodic latency spikes under SR.
We traced this to placement instability: with low per-client rates, the
invocation count in each sliding window varies randomly, causing the estimated
benefit of marginal SR followers to oscillate around the $\alpha \cdot
  \text{size}_c$ threshold. Since GV rebuilds its placement from scratch at each
iteration, these followers are added and omitted in alternation.

\subsection{Adaptivity to Mobile Clients}
\label{sec:eval-adaptivity}

To answer RQ4, we reuse the deployment, workload, and policy configuration of
\s{eval-decentralized} with mobile clients replacing the static ones.
We introduce a cyclic mobility pattern (round-robin) to model typical user
routines, such as commuting between a fixed set of locations. For each client,
we randomly assign a predefined sequence of $N$ waypoints (with $N$ chosen
uniformly between 2 and 4) from all available nodes within its native tier
(edge or fog). The client periodically relocates to the next node in its
sequence. The movement interval is fixed at 30\,s, comparable to the
sliding-window size $W = 32$\,s, so that each re-planning cycle is likely to
span a client relocation.
Mobility is the distinguishing factor of this experiment. Centralized policies
observe demand through a sliding window of the most recent $W$ seconds; when
clients move within a window, the rate estimates blend old and new locations,
and the resulting placement reflects a demand distribution that no longer
exists. AV, by contrast, reacts to each invocation individually and therefore
adapts continuously. We expect this structural advantage to manifest as a
lower and more stable latency under mobility.

\fakeparagraph{Results}
\fig{latency-combined} shows the median latency over time under round-robin
mobility.
BLP exhibits recurring latency spikes in all three scenarios. Each spike
corresponds to a period during which the active plan is stale: the solver
produces a placement optimized for a demand distribution that has already
shifted. Under SR the spikes are smaller in amplitude, because the co-location
constraint limits how far the placement can deviate from the true optimum even
with outdated rate estimates.
GV tracks a smoother profile than BLP, as both share the same sliding-window
estimation and gather-scatter overhead but GV solves instantaneously, avoiding
the stale-plan intervals that cause BLP spikes. Under ER, GV latency is
comparable to BLP~0.2s, while BLP~10s suffers from larger spikes due to its
longer re-planning cycle leaving stale plans in place across client moves.
Under MIX, the contrast is sharper: GV maintains a relatively stable latency,
whereas both BLP variants exhibit recurring spikes.
Under SR, however, GV performs worse than the other centralized policies on
average, with a persistently elevated baseline. This may indicate that the
greedy SR placement, which fixes the leader based on the initial
write-pressure snapshot, does not adapt well when the writer population shifts
due to mobility; the BLP, by contrast, can re-elect the leader at each
iteration.
AV achieves the lowest and most stable latency across all three consistency
mixes. Under ER and MIX, AV settles after a brief initial transient and
maintains a flat profile throughout. This is a direct consequence of its
per-invocation reactivity: when a client moves to a new node, the always-copy
mechanism begins pulling replicas toward the new location from the first
invocation, and the hop-by-hop scheduling routes requests through the
hierarchy without waiting for a global re-plan. Under SR, AV converges equally
quickly, as the force model routes writers to the leader and readers to the
nearest follower independently of client movement.

\subsection{Discussion}
\label{sec:eval-discussion}

Taken together, the four RQs delineate a clear trade-off between optimality,
scalability, and runtime adaptivity. BLP remains valuable as a reference
point, but only within a limited problem size. GV preserves almost all of its
placement quality while extending the tractable regime by orders of magnitude,
making a centralized global-view heuristic the most credible option when the
workload can be summarized reliably.
AV is not designed to match the placement quality of BLP or GV on a static
snapshot; it reacts to each invocation as it arrives. Under stable demand the
two approaches yield comparable results; under mobility the gap is
substantial, as centralized policies rely on rate estimates that grow stale
within a single re-planning window, while AV adjusts placement continuously.
Finally, the latency degradation of the CO and CD baselines reinforces our
main claim: pushing computation to the edge without a coupled,
consistency-aware data placement policy misses the main source of performance
improvement. %
\section{Conclusions}
\label{sec:conclusions}

This paper addressed the joint problem of function scheduling and data
placement in the edge-cloud continuum under two consistency models, SR and ER.
We proposed three approaches with progressively narrower information scope: a
BLP that computes the optimal placement given a complete system snapshot, a GV
greedy heuristic that operates on the same snapshot but at a fraction of the
computational cost, and an AV decentralized protocol in which each node makes
placement and scheduling decisions based on locally observed demand.
Overall, the results show that stateful serverless on the edge-cloud continuum
requires a joint treatment of function scheduling and data placement, and that
the most suitable decision mechanism depends on workload stability. A
centralized global view is effective when demand can be summarized reliably,
whereas under runtime shifts a decentralized policy such as AV can better
translate local information into timely adaptation.

Several directions remain open.
First, our model assumes static knowledge of function-to-collection access
patterns; extending it to discover or learn these dependencies at runtime
would broaden the applicability of the framework.
Second, scenarios with heavily loaded edge nodes may benefit from
incorporating compute budgets into the placement decision; how to characterize
function memory footprints accurately enough to make such budgets actionable
remains an open question.
Finally, we plan to integrate the AV protocol into a serverless platform,
bridging the gap between the simulated evaluation presented here and an
operational deployment in the edge-cloud continuum. 
\printcredits

\section*{Data availability}
Code, experiment configurations, raw results, and the notebooks that regenerate
all figures in this paper are openly available~\citep{artifact}.

\section*{Declaration of competing interest}
The authors declare that they have no known competing financial interests or
personal relationships that could have appeared to influence the work reported
in this paper.

\section*{Funding}
This research did not receive any specific grant from funding agencies in the
public, commercial, or not-for-profit sectors.

\section*{Declaration of generative AI and AI-assisted technologies in the manuscript preparation process}

During the preparation of this work, the authors used generative AI in order
to refine language editing and support software development;  all content was
reviewed and verified by the authors, who assume full responsibility for the
manuscript.

\bibliographystyle{elsarticle-num}

\end{document}